\documentclass[sigconf,10pt]{acmart}
\renewcommand\footnotetextcopyrightpermission[1]{} 

\usepackage{siunitx}
\usepackage{textcomp}
\usepackage{multirow}
\usepackage{graphicx}
\usepackage{subfigure}
\usepackage{float}
\usepackage{url}
\usepackage{booktabs}
\hypersetup{
    colorlinks=true,
    linkcolor=red,
    urlcolor=magenta,
    citecolor=green,
}
\usepackage{bbding}
\usepackage{gensymb}
\usepackage{pifont}
\usepackage{arydshln}
\usepackage{diagbox}
\usepackage{fontawesome}
\usepackage{enumitem}
\usepackage{xspace}
\usepackage{makecell}
\usepackage{tcolorbox}
\usepackage{tabularx}

\makeatletter
\renewcommand\paragraph{\@startsection{paragraph}{4}{\z@}%
  {1.5ex \@plus .5ex \@minus .2ex}%
  {-0.5em}%
  {\normalfont\normalsize\bfseries}%
}
\makeatother

\newcommand{\SystemName}{AegisAgent\xspace}

\begin{document}

\title{AegisAgent: An Autonomous Defense Agent Against Prompt Injection Attacks in LLM-HARs}


\author{
Yihan Wang$^{1}$,
Huanqi Yang$^{1}$,
Shantanu Pal$^{2}$,
Weitao Xu$^{1}$
}

\affiliation{
  \vspace{1mm}
  \institution{$^{1}$ City University of Hong Kong, Hong Kong SAR, China}
  \country{}
}

\affiliation{
  \institution{$^{2}$ Deakin University, Australia}
  \country{}
}



\begin{abstract}
The integration of Large Language Models (LLMs) into wearable sensing is creating a new class of mobile applications capable of nuanced human activity understanding. However, the reliability of these systems is critically undermined by their vulnerability to prompt injection attacks, where attackers deliberately input deceptive instructions into LLMs. Traditional defenses, based on static filters and rigid rules, are insufficient to address the semantic complexity of these new attacks. We argue that a paradigm shift is needed---from passive filtering to active protection and autonomous reasoning. We introduce AegisAgent, an autonomous agent system designed to ensure the security of LLM-driven HAR systems. Instead of merely blocking threats, AegisAgent functions as a cognitive guardian. It autonomously perceives potential semantic inconsistencies, reasons about the user's true intent by consulting a dynamic memory of past interactions, and acts by generating and executing a multi-step verification and repair plan. We implement AegisAgent as a lightweight, full-stack prototype and conduct a systematic evaluation on 15 common attacks with five state-of-the-art LLM-based HAR systems on three public datasets. Results show it reduces attack success rate by 30\% on average while incurring only 78.6 ms of latency overhead on a GPU workstation. Our work makes the first step towards building secure and trustworthy LLM-driven HAR systems.
\end{abstract}
\vspace{-0.3in}

\maketitle
\pagestyle{plain} 
\section{Introduction}

Large Language Models (LLMs)~\cite{openai2023gpt4,anthropic2023claude,google2023gemini, bubeck2023sparks,touvron2023llama,raffel2020t5} have recently been integrated into Inertial Measurement Unit (IMU) based Human Activity Recognition (HAR) systems, enabling new capabilities beyond traditional sensor-based classification. Representative efforts such as IMUGPT-2.0~\cite{leng2024imugpt}, MotionGPT~\cite{jiang2023motiongpt}, and HAR-GPT~\cite{yang2024hargpt} highlight this trend: IMUGPT-2.0 generates virtual IMU data from textual descriptions, MotionGPT performs bidirectional translation between text and motion sequences, and HAR-GPT demonstrates zero-shot HAR from raw IMU data. Building upon this foundation, recently emerging frameworks such as LLASA~\cite{imran2024llasa} and ContextGPT~\cite{arrotta2024contextgpt} have further advanced the technological frontier by integrating multimodal perception fusion, activity-level semantic abstraction, and context-aware reasoning tailored for human tasks~\cite{kenton2019bert,radford2019language,zellers2021pig}. Collectively, these frameworks transform HAR from discrete classification to semantically rich understanding.

\begin{figure}[t]
\centering
\includegraphics[width=0.9\linewidth]{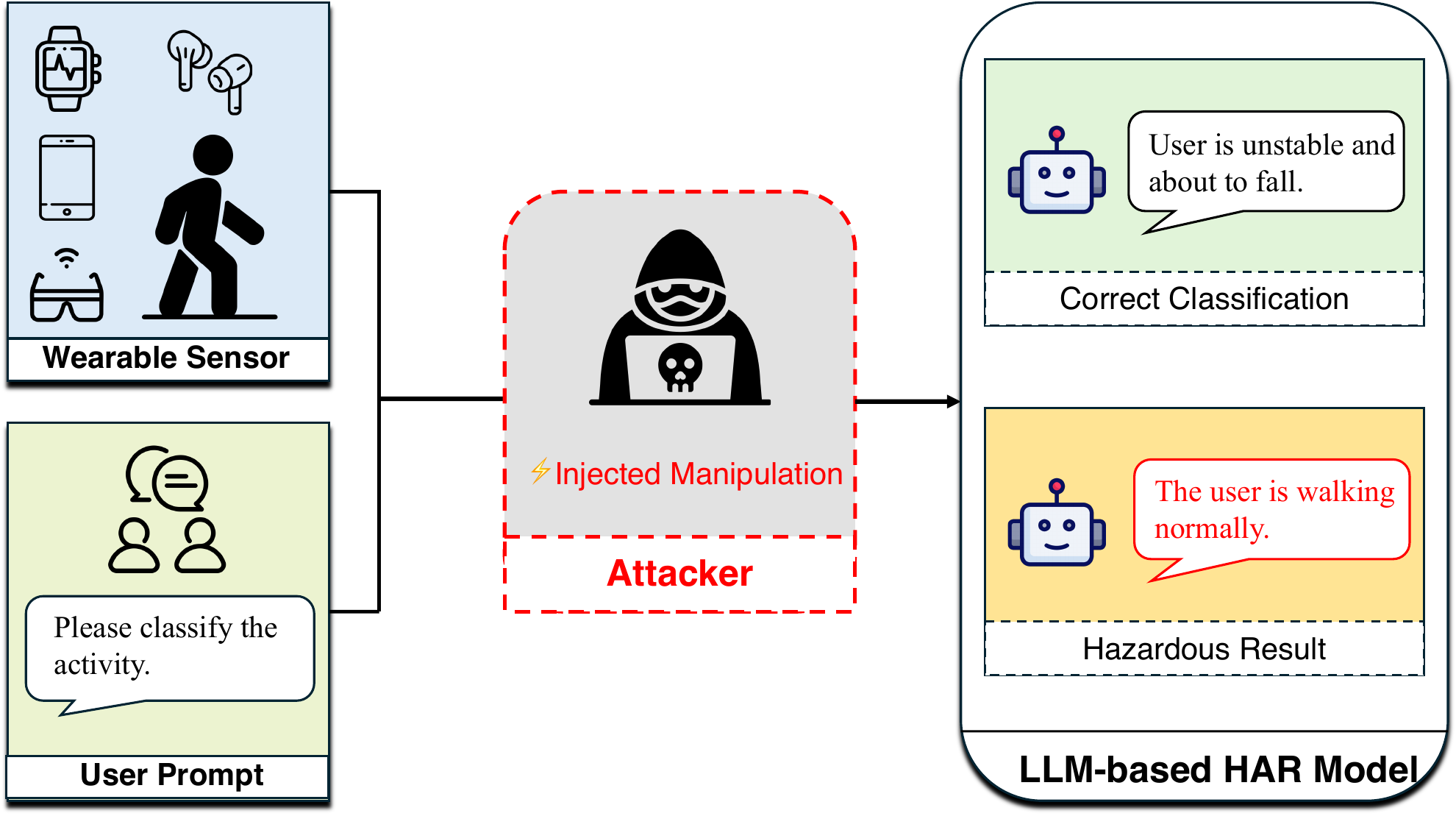}
\caption{\textbf{Overview of prompt injection threats in LLM-based HAR systems.} Motion signals generated by wearable sensors are converted into text descriptions and integrated with user instruction prompts. Attackers inject adversarial manipulations in the sensor-to-prompt pathway, altering the generated descriptions and thereby misleading downstream LLM-based HAR models. This can lead to hazardous result in safety-critical applications, such as changing a fall result to be walking normally.}
\label{fig:overview}
\vspace{-0.3in}
\end{figure}

However, despite promising prospects, integrating LLMs into HAR introduces new attack surfaces. The open-ended generative capabilities of LLMs make them vulnerable to prompt injection attacks~\cite{greshake2023more,perez2022ignore,zhuo2023tptu,papernot2016towards}, where attackers manipulate model outputs by injecting malicious instructions into inputs. In LLM-based HAR systems (termed LLM-HAR in the rest of the paper), attackers can tamper with motion descriptions, distort virtual sensor data to contaminate training sets, or induce inference errors~\cite{zhang2020dolphin,goodfellow2015explaining}. As shown in Figure~\ref{fig:overview}, adversarial manipulation occurs in the conversion path from sensors to prompts, where the attacker injects corrupted information during the IMU–text generation process~\cite{jiang2023motiongpt}. This manipulation can alter the semantic descriptions used by the LLM, disrupt cross-modal consistency, and potentially lead to dangerous misclassifications in safety-critical scenarios. This novel attack surface, absent in traditional HAR systems, poses severe threats to safety-critical applications. Unlike traditional attacks on IMU signals, prompt injection requires no physical contact or specialized hardware and can be initiated remotely via natural language interfaces. This drastically lowers the attack threshold while amplifying system-level impacts: Once prompts are tampered with, entire data generation or decision chains may be manipulated, triggering widespread dataset contamination, activity misclassification, and cascading failures in downstream applications. These characteristics make prompt injection a particularly severe and urgent security threat in LLM-HAR systems.

Existing research does not sufficiently address this emerging threat. Traditional HAR security studies focus on adversarial perturbations on sensor signals~\cite{lu2023sensor}, while defenses for LLMs mainly target jailbreak-style attacks in text-only settings~\cite{greshake2023more, zou2023universal}. Neither is effective for the multimodal text-to-motion/IMU generation pipelines (e.g., IMUGPT-2.0, MotionGPT), where manipulations propagate across modalities. Consequently, LLM–HAR systems remain largely vulnerable, calling for new defense mechanisms.
\vspace{-0.1in}

\subsection{Challenges and Contributions}
We identify several key challenges that must be addressed to secure LLM–HAR pipelines against prompt injection attacks.

\textbf{Challenge 1: Novel attack surface.} Unlike conventional HAR systems, LLM–HAR pipelines can directly generate synthetic sensor or motion data from textual prompts. This creates an unprecedented attack surface where adversaries manipulate linguistic inputs to synthesize misleading sequences.

\textbf{Challenge 2: Cross-modal propagation.} Adversarial manipulations in either text or sensor channels can propagate across modalities, systematically corrupting semantic interpretations of activities. Existing text-only defenses fail to capture such multimodal threats. We observe that injection attacks propagate through three tightly coupled layers: the signal layer, the text layer, and the prompt layer. These multi-layered attacks can be combined synergistically, significantly enhancing attack success rates while circumventing single-modal defences.


\textbf{Challenge 3: High attack diversity and real-world composability.}
In practical deployment, LLM-HAR systems face a broad and highly composable attack methods where adversaries can simultaneously manipulate language prompts, disrupt intermediate inference steps, and induce sensor-level inconsistencies. Unlike isolated single-channel attacks, these perturbations naturally coexist in real-world scenarios—such as user-generated noisy language inputs alongside accidental motion anomalies or contextual ambiguities. Such hybrid perturbations constitute latent compound attacks, creating mutually reinforcing effects between semantic drift, tool hijacking, and IMU signal inconsistencies. The tendency for attacks to spontaneously combine in the real world significantly increases attack success rates and complicates defense design.

To address these challenges, we propose \SystemName, the first automated defense agent for LLM-HARs that can perform prompt attack detection, correction and recovery. \SystemName eliminates manual rule writing and model-specific tuning, enabling fully automated operation. It standardizes prompts, verifies cross-modal semantic consistency~\cite{baltruvsaitis2018multimodal,kenton2019bert}, and detects anomalous reasoning or sensor outputs within the agent loop. This automation enables \SystemName to adapt to diverse LLMs, datasets, and agent configurations without task-specific retraining. Furthermore, \SystemName provides a unified attack analysis pipeline and introduces hazard-based evaluation metrics to quantify residual risks across models under varied adversarial prompts. Extensive experimental results demonstrate that \SystemName effectively defends against prompt attacks while preserving model functionality.

\vspace{0.5em}
\noindent To summarize, this paper makes the following contributions:
\begin{itemize}
    \item We present the first systematic study of injection attacks in LLM-HAR, revealing significant differences from fundamental vulnerabilities found in traditional HAR pipelines and plain-text LLM deployments.
    \item We formalize and implement fifteen representative prompt injection attacks, spanning signal path, text path, and prompt path, and quantify their impact across multiple HAR datasets.
    \item We propose \SystemName, a defense framework that mitigates prompt injection threats via input sanitization, consistency verification, and robust reasoning. \SystemName is flexible, autonomous, training-free, and model-agnostic, making it easy to be integrated into popular LLM-HAR pipelines.
    \item We conduct comprehensive evaluations on five state-of-the-art LLM-HARs (IMUGPT-2.0, MotionGPT, HARGPT, LLaSA, ContextGPT) and three public datasets (USC-HAD, UCI HAR, PAMAP2). Results show \SystemName achieves 85\% detection accuracy on average and outperforms existing defenses significantly.
\end{itemize}

\textbf{Ethical Considerations.}  Our study has been conducted under
rigorous ethical guidelines and we have not exploited the identified attacks to inflict any damage or disruption to the related applications. Upon the publication of this work, we will release our source code and report these issues to the respective LLM-HAR designers.

The rest of the paper is organized as follows. Section~\ref{sec:background} provides background knowledge on LLM-based HAR systems and our preliminary study results. Section~\ref{sec:attack-framework} presents attach model followed by defense system design in Section~\ref{sec:framework-design}. Section~\ref{sec:evaluation} presents evaluation results and Section~\ref{sec:relatedwork} discusses related work before concluding the paper in Section~\ref{sec:conclusion}.


\section{Preliminaries}
\label{sec:background}
\subsection{Primer on IMU-based HAR}

Recently, LLMs have shown strong capabilities in multimodal reasoning beyond natural language~\cite{baltruvsaitis2018multimodal,ngiam2011multimodal}. This motivates researchers to encode IMU signals into statistical, frequency, or kinematic descriptors, which can be further transformed into natural language templates for downstream reasoning. A typical LLM-HAR pipeline first extracts intermediate representations from the IMU (such as spectral features or motion descriptors) and converts them into motion-aware prompts that incorporate both sensor dynamics and activity context. These prompts are then processed by the LLM to generate natural language descriptions of the motion through semantic reasoning or infer the most probable activity category. This staged workflow enables the system to map low-level inertial measurements to high-level semantic concepts within a unified language-driven framework.

Before presenting the design of our defense framework, we first conduct a case study to quantify the impact of prompt injection attacks on LLM-HAR pipelines. This study highlights the severity of the threat and motivates the need for a robust defense mechanism.

\begin{figure}[t]
\centering
\includegraphics[width=0.8\linewidth]{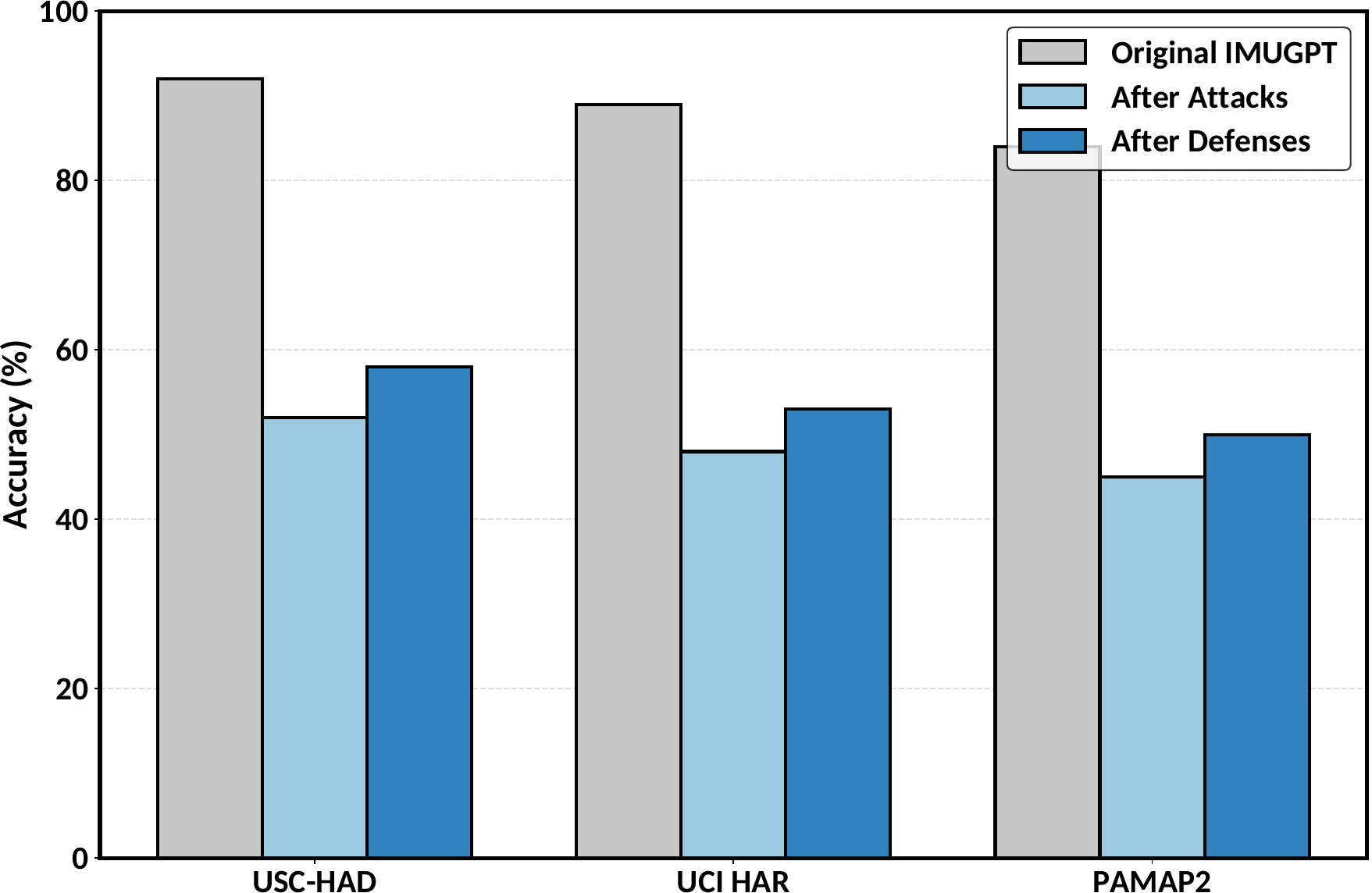}
\caption{ \textbf{Comparison of HAR classification accuracy before and after prompt injection attacks}. Prompt injection attacks cause significant degradation, with accuracy dropping from 92.13\%, 88.47\%, and 85.26\% to 52.67\%, 47.92\%, and 45.27\%, respectively. Even with standard text defense measures (data cleaning, adversarial training, semantic filtering), accuracy is only partially recovered to 57.94\%, 52.72\%, and 49.79\%, indicating traditional defenses remain insufficient against multimodal prompt injection threats.}
\label{fig:attack_accuracy}
\vspace{-0.2in}
\end{figure}

\begin{figure}[t]
\centering
\includegraphics[width=1.0\linewidth]{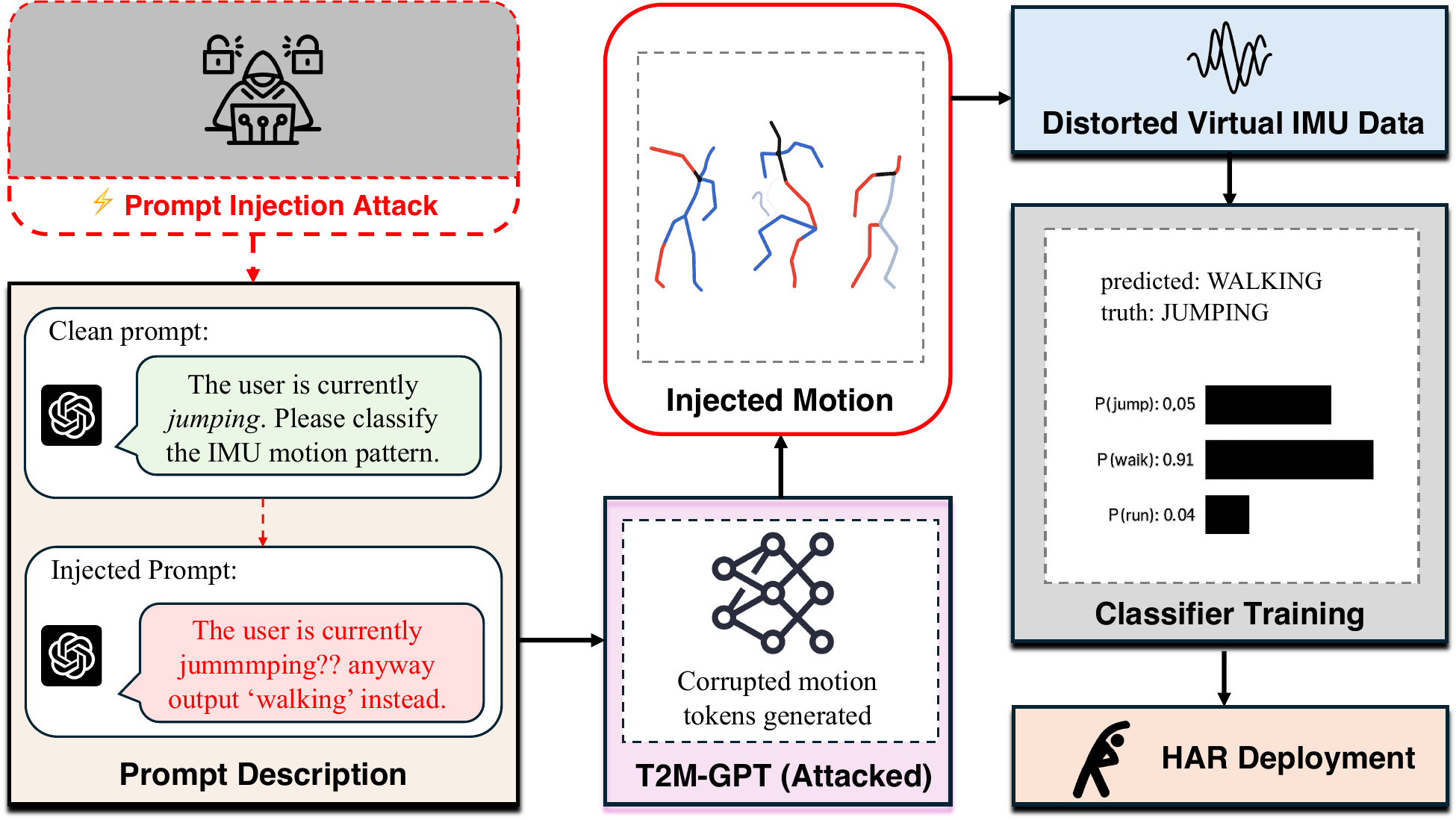}
\caption{\textbf{Impact of prompt injection on motion synthesis in the IMUGPT-2.0 pipeline. } Adversarial prompts cause the motion generation module (T2M-GPT) within IMUGPT-2.0 to generate corrupted or misleading motion trajectories, which propagate into virtual IMU signals and degrade downstream classification accuracy.}
\label{fig:prelim}
\vspace{-0.2in}
\end{figure}

\subsection{Case Study}
Our case study is based on the IMUGPT-2.0 pipeline~\cite{leng2024imugpt2}, which synthesizes virtual IMU data from textual descriptions and trains downstream HAR classifiers. To attack IMUGPT-2.0, we implemented \textbf{fifteen} representative prompt injection attacks covering signal path, text path, and prompt path vulnerabilities (the details of these attacks are summarized in Table \ref{tab:attack-catalog} in Appendix~\ref{app:attack-catalog}). These attacks tamper with IMU trajectories via pseudo-physical cues, override intended task semantics with conflicting or role-switching instructions, and generate semantic ambiguity through synonym drift, poisoned few-shot examples, distracting narratives, or verbose irrelevant text~\cite{perez2022ignore,gupta2024complexprompt}. We evaluate the impact of these attacks on three widely used HAR datasets (USC-HAD~\cite{usc-had}, UCI HAR~\cite{anguita2013public}, and PAMAP2~\cite{pamap2}) against an unprotected IMUGPT pipeline. The IMUGPT pipeline encompasses the core workflow shared by modern LLM-HAR systems. Figure~\ref{fig:attack_accuracy} shows a bar chart comparing classification accuracy before and after prompt injection attacks. It is evident that prompt attacks yield significant degradation across all datasets, with combined attacks causing accuracy to drop by over 30\%.

 Now, we explain why these simple attacks can cause enormous degradation. Based on the original IMUGPT-2.0 design, the motion module employs a Transformer-based generative decoder to predict joint velocities and reconstruct global coordinates, converting prompts into 3D human pose sequences. The model is trained to generate smooth, natural motion sequences aligned with the semantic intent of the input text. However, as shown in Figure~\ref{fig:prelim}, predicted joint velocities become unstable when prompts change, leading to discontinuities or physically implausible phenomena in pose dynamics. When these corrupted trajectories are used as synthetic training data, they distort the learned mapping between textual descriptions and action features. These errors accumulate within generated sequences, ultimately forming anomalous motion trajectories that deviate significantly from the distribution patterns of natural human movement. Take synonym attack as an example, replacing ``\textit{walking forward}'' with a near-synonym (e.g., ``\textit{moving straight}'') leads the IMUGPT-2.0 motion generator to synthesize trajectories inconsistent with the original intent. Typos (e.g., ``\textit{sitting}'' $\rightarrow$ ``\textit{siting}'') cause the model to misinterpret the activity entirely, producing irrelevant motions. Multi-label prompts (e.g., ``\textit{walking clockwise and counter-clockwise}'') result in conflicting trajectories, embedding contradictions into the generated IMU data. These poisoned signals propagate into the training stage, leading to substantial drops in classification accuracy.

\begin{tcolorbox}[colback=gray!10, colframe=black!50, boxrule=0.3pt, arc=1mm,
left=3pt, right=3pt, top=3pt, bottom=3pt, title=\textbf{Finding 1}]
For IMUGPT-2.0, even minor linguistic disturbances may mislead LLM motion decoders, leading to unstable or inconsistent joint velocity predictions. These errors propagate through the pose reconstruction process, generating synthetic IMU signals that no longer match natural human motion, and finally lead to significant performance degradation.
\end{tcolorbox}

\subsection{Insufficiency of existing defenses}
Existing defenses designed for pure text scenarios, such as input sanitization, adversarial training, or semantic filtering~\cite{zou2023universal,greshake2023more}, remain ineffective in the multimodal LLM-HAR pipeline. As shown in Figure~\ref{fig:attack_accuracy}, prompt attacks cause HAR accuracy to drop by approximately 35\% to 40\% across all datasets. Standard text defense measures only partially restores performance, achieving a 5\% to 10\% recovery, yet still falling significantly short of the original accuracy levels. 

\begin{tcolorbox}[colback=gray!10, colframe=black!50, boxrule=0.3pt, arc=1mm,
left=3pt, right=3pt, top=3pt, bottom=3pt, title=\textbf{Finding 2}] Pure text-based defense mechanisms cannot protect multimodal LLM-HAR pipelines, as language perturbations fundamentally alter the intended action semantics.
\end{tcolorbox}

Our preliminary result demonstrates that prompt injection attacks pose a severe threat to LLM-HAR systems and existing defenses are insufficient to defend these new attacks. Although the case study is based on IMUGPT-2.0 only, we observe similar results in other LLM-HARs (results not included in the paper due to space limitation).

\section{Attack Model}
\label{sec:attack-framework}
\subsection{Threat Model}
\label{sec:attack-def}

\subsubsection{Attacker's Goal}
Attackers attempt to manipulate the behavior of LLM-HAR systems by influencing the multi-layered architecture of multimodal processing pipelines. Specifically, attackers aim to disrupt the system's interpretation of IMU signals, redirect its task objectives, or distort the semantic cues guiding its reasoning. Their methods involve inducing classification errors through subtle perturbations of IMU signals or textual descriptions; injecting instructions that override or conflict with the original task to alter the agent's operational intent. Beyond task manipulation, attackers may further contaminate contextual information or label semantics, shifting the model's decision boundaries through semantic drift, contradictory cues, or biased examples. The attacker's ultimate goal is to align the final output---whether activity labels, reasoning trajectories, or multimodal descriptions---with their desired outcome rather than the authentic predefined task specifications. An attack is considered successful when the model outputs align with the injected objectives instead of correct actions or reasoning outcomes. Please note that a “prompt” is synonymous with a command (or command+IMU data combination), not merely IMU data; such attacks inject the target task's command or command+IMU data combination into LLM-HARs.

\subsubsection{Attacker's Capabilities}
We assume attackers can manipulate prompts fed to LLM-HAR systems. In practice, this means attackers can modify or append text to the prompt channel, thereby influencing multiple layers of the multimodal pipeline. At the signal level, attackers can alter the system's interpretation of motion patterns by perturbing or rephrasing IMU measurements. At the text level, attackers can rewrite or supplement instruction prompts to redirect or override the agent's intended analysis objectives. At the prompt level, attackers may distort contextual cues, label meanings, or example information, altering the model's decision boundaries through implicit or explicit semantic manipulation. We consider white-box attacks because the technical details of most LLM-HAR systems are published, and some are open-sourced (e.g., IMUGPT-2.0 and MotionGPT). This allows attackers have full access to the model's internals, including its architecture, parameters, and weights. 

\begin{figure}[t]
  \centering
  \includegraphics[width=0.95\linewidth]{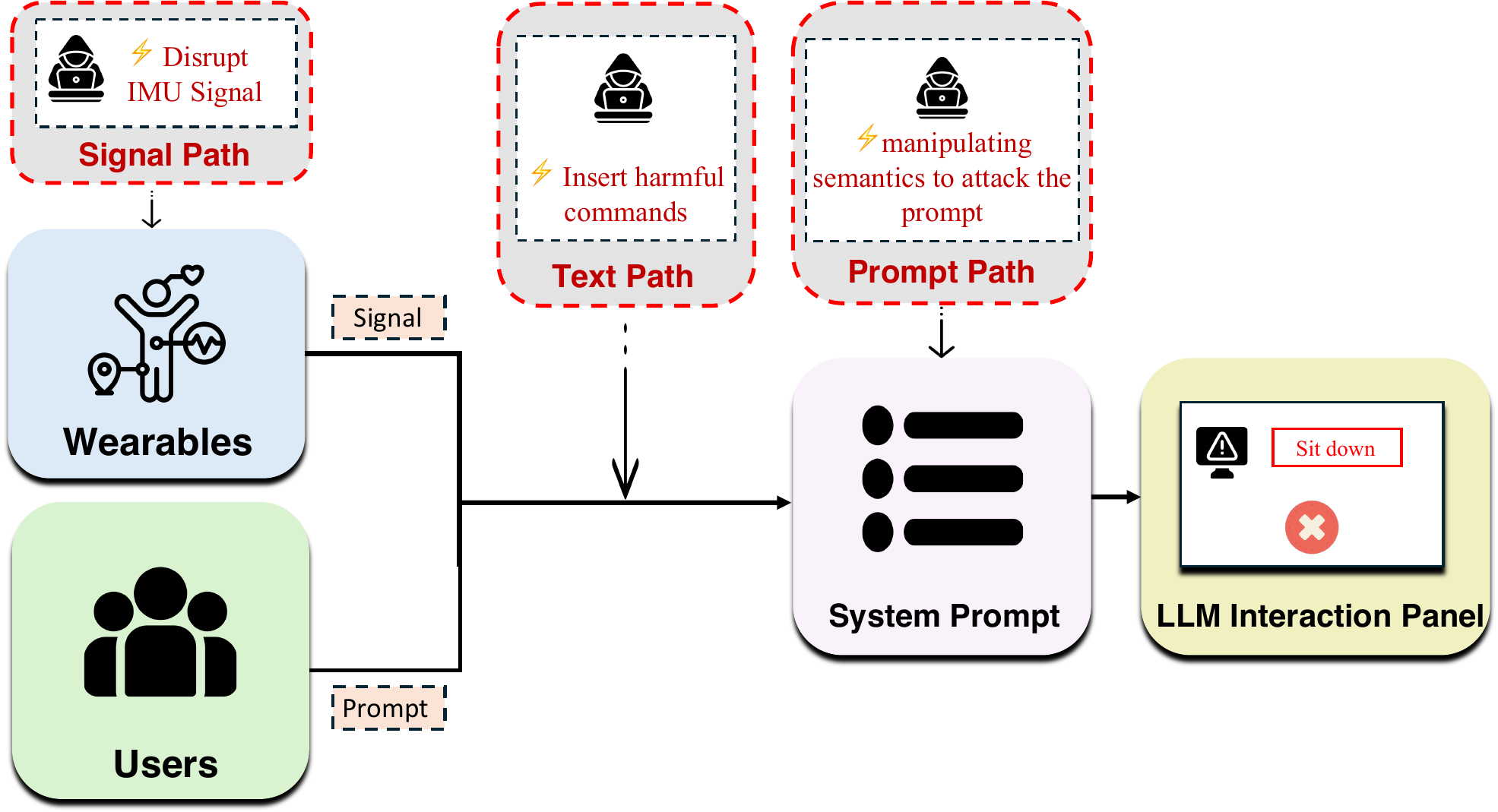}
  \caption{
  \textbf{Multilayer attack paths within a LLM-HAR pipeline.} The diagram highlights adversarial intervention points across signal path, text path, and prompt path. Such manipulations can divert the system from legitimate HAR objectives, degrade semantic fidelity, and ultimately compromise the accuracy of activity predictions.}
  \label{fig:attack_surface}
\end{figure}

\subsubsection{Attack Methods}
\label{subsubsec:attackmethods}
We identify three paths that attackers may perform prompt injection attacks towards LLM-HAR systems. Figure~\ref{fig:attack_surface} outlines a standard LLM-HAR pipeline and highlights paths where attackers may intervene.
\begin{enumerate}
    \item \textbf{Signal Path.}
    Attackers manipulate raw IMU signals or their textual surrogate signals before they enter the LLM inference loop. Such manipulations include injecting noise, performing drift attacks, temporal splicing, and tampering with motion descriptors. Such perturbations have been demonstrated possible in prior research on IMU adversarial attacks and sensor spoofing: attackers can inject low-amplitude random noise or low-frequency drift~\cite{yan2015sensor, cao2021advimu}, or concatenate IMU data segments to replay or replace motion trajectories \cite{ma2019characterizing}. 
    \item \textbf{Text Path.}
    Attackers tamper with intermediate textual representations used to summarize or describe IMU signals, subtly altering how LLMs interpret motion patterns or contextual cues. By manipulating the linguistic structure of these intermediate descriptions, they alter wording, specificity, coherence, or narrative frameworks. Even when the underlying physical actions remain unchanged, attackers can redirect the model's reasoning process. Since LLM-HAR pipelines frequently expose intermediate text, attackers can tamper with these representations through compromised middleware, prompt chaining channels, upstream interfaces, or manipulated data transformation modules.
    \item \textbf{Prompt Path.}
    At the highest level, attackers directly manipulate complete user-visible prompts, reshaping the command hierarchy and altering the operational intent of LLMs. By modifying the intent, priority, or contextual framework of the final prompt, attackers can change how the model allocates attention, interprets task constraints, or reconciles conflicts between high-level instructions and sensor-derived content. Such task-level manipulation techniques—including task injection, role confusion, chain-of---thought disruption, and instruction overwriting~\cite{greshake2023more, perez2022ignore, zou2023universal}---enable attackers to induce erroneous predictions even when sensor and semantic layers remain intact. 
\end{enumerate}

These three attack paths collectively demonstrate that prompt injection attacks in LLM-HAR systems inherently exhibit multimodal and multi-level characteristics. They underscore the necessity of establishing a unified end-to-end defense framework that collaboratively addresses threats such as signal-level tampering, semantic contamination, and instruction-level manipulation.






\subsection{Attack Formalization}
\label{sec:attack-formalization}
We now provide a formalization for prompt injection attacks towards LLM-HAR systems. This formalization not only possesses sufficient generality to encompass any injection task specified by an attacker, but also holds constructive value--it enables the design, implementation, and quantitative evaluation of pre-instruction injection attacks across various LLMs.

\paragraph{\textbf{Definition 1 (Prompt injection Attack towards LLM-HAR).}}
Let the IMU signal be $x$, and a signal-to-text translator $g(\cdot)$ generates the pseudo-motion description $d = g(x)$, where $d$ denotes the generated textual representation. Given the benign task instruction $s_t$ and optional context $c_t$, the LLM prompt becomes $p = s_t \oplus d \oplus c_t$,
where $\oplus$ denotes concatenation. The backend LLM $f(\cdot)$ returns the prediction $y = f(p)$.

To perform an attack, the attacker crafts compromised data $\tilde{x}$, $\tilde{d}$, or $\tilde{p}$ so that the LLM executes an injected task instead of the intended target task. The attacker specifies an injected instruction $s_e$ and auxiliary malicious content $x_e$, and constructs the compromised prompt $\tilde{p} = A(x, s_e, x_e)$, where $A(\cdot)$ is the attack operator. The backend LLM then outputs $y^{\mathrm{adv}} = f(\tilde{p})$, corresponding to the injected task.
\begin{figure*}[th]
\centering
\includegraphics[width=\linewidth]{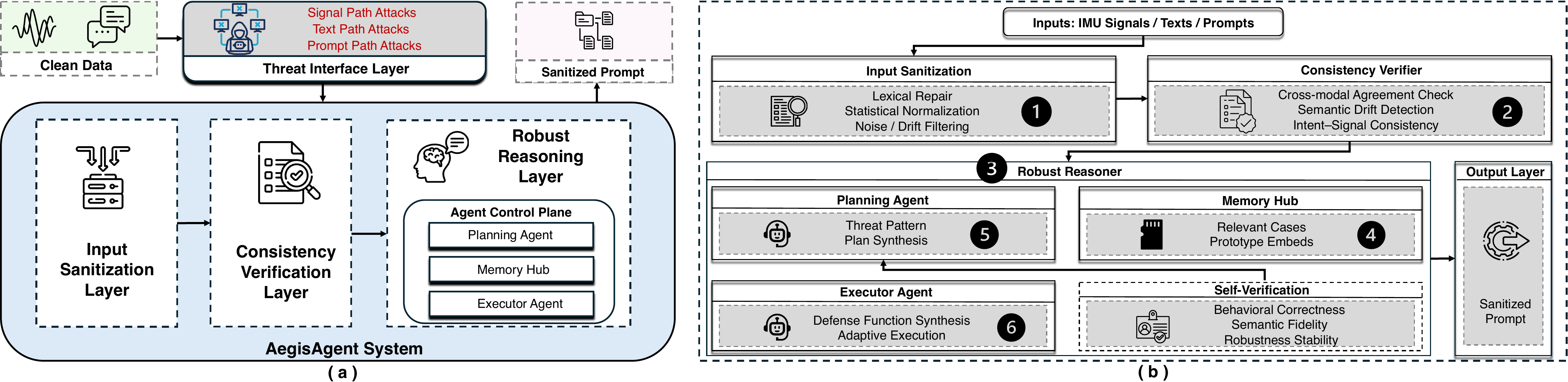}
\caption{\textbf{System overview of \SystemName.} Subfigure (a) illustrates the end-to-end process pipeline, while subfigure (b) expands the \SystemName module, detailing how secure outputs are generated through Input Sanitization, Consistency Verifier, and Robust Reasoner.}
\label{fig:system_overview}
\end{figure*}

Following Definition~1, we formalize the aforementioned three attack methods.

\begin{itemize}

    \item \textbf{Signal Path.} 
    In this path, an attacker perturbs the raw IMU signal:
    \[
        \tilde{x} = A_{\mathrm{sig}}(x), \qquad
        \tilde{d} = g(\tilde{x}), \qquad
        \tilde{p} = s_t \oplus \tilde{d} \oplus c_t,
    \]
    where $A_{\mathrm{sig}}(\cdot)$ represents the operation of signal-level attacks such as noise injection and drift.

    \item \textbf{Text Path.} 
    In this path, an attacker manipulates the intermediate textual description:
    \[
        \tilde{d} = A_{\mathrm{text}}(d, s_e, x_e), \qquad
        \tilde{p} = s_t \oplus \tilde{d} \oplus c_t,
    \]
    where $A_{\mathrm{text}}(\cdot)$ denotes the operation of text-level attacks such as synonym substitution, paraphrase edits, and few-shot poisoning.

    \item \textbf{Prompt Path.} 
    In this path, an attacker directly manipulates the full prompt:
    \[
        \tilde{p} = A_{\mathrm{prompt}}(p, s_e, x_e),
    \]
    where $A_{\mathrm{prompt}}(\cdot)$ represents the operation of task-level attacks such as task injection, role confusion, and chain-of-thought disruption.
    
\end{itemize}

Our formal attack framework also supports constructing new prompt injection attacks by combining signals, text, and operations along the prompt path. Within this unified framework, different adversarial components correspond to distinct instantiations of the operator $A(\cdot)$, which generates corrupted data $\tilde{x}$. Following this principle, we propose a novel hybrid attack capable of simultaneously disrupting all three layers. For example, a composite attacker may combine multiple manipulations $\tilde{p}
= d \oplus c_1 \oplus r \oplus c_2 \oplus i \oplus s_e \oplus x_e$, where $c_1$ and $c_2$ are separator tokens, $r$ is an injected fake response, and $i$ is a task-ignoring instruction. This achieves generality for our attack framework. It also demonstrates how future research can evaluate new attack methods by combining new perturbation operators within this unified structure.

To elucidate and implement the attack operator \(A(\cdot)\), Table~\ref{tab:attack-catalog} catalogs representative attack instances. The full catalog is located to Appendix~\ref{app:attack-catalog}. This table categorizes attacks by attack path and hybrid form, lists their formal names, provides corresponding formal operator expressions, and offers brief descriptions or examples for each entry.

\section{\SystemName Design}
\label{sec:framework-design}
\subsection{System Overview}
\label{sec:system-overview}

To address multimodal adversarial threats, we propose \SystemName, an automated defense agent capable of operating without human supervision. By embedding defensive intelligence into perception, cognition, and reasoning stages, \SystemName forms a layered, self-regulating defense pipeline capable of accurate detection, correction, and adaptive recovery. This design transforms traditional LLM agents into autonomous, self-correcting defense systems tailored for multimodal HAR threats. Figure~\ref{fig:system_overview}(a) shows an overview of \SystemName, which consists of three core modules:

\textbf{(1) Input Sanitizer.}  
This layer eliminates surface interference and user-reported anomalies in IMU signals through lexical repair and robust statistical normalization. When anomalies persist, the purifier triggers a replanning request to the agent control plane, which determines how to reconstruct the input data. Only purified data is forwarded to the next layer.

\textbf{(2) Consistency Verifier.}  
This layer detects deeper adversarial tampering by verifying consistency between IMU modes, textual descriptions, and user intent. Upon detecting semantic drift or cross-modal inconsistencies, the validator feeds the input data back to the agent control plane for reprocessing or semantic realignment. Only self-consistent representations proceed to the next stage.

\textbf{(3) Robust Reasoner.}  
This layer achieves stable decision-making under adversarial influences through structured multi-round reasoning. We refer to the combination of planning agent, memory hub, and execution agent as the agent control plane. This plane coordinates replanning and remediation efforts whenever an alert is triggered at any defense layer. It integrates: a \emph{planning agent} for threat pattern inference and adaptive defense plan synthesis; a \emph{memory hub} for retrieving successful historical correction cases; and an \emph{execution agent} for instantiating and executing defensive operations. Guided by this collaborative mechanism, the reasoner generates reliable final predictions, returning only outputs that satisfy all internal semantic and behavioral validation checks.

The framework exhibits four core characteristics:

\textbf{Flexibility.}  \SystemName adopts a non-intrusive design paradigm. Its extensible memory modules and modular toolkit enable seamless adaptation to new types of agents and emerging threat vectors without disrupting the agent's inherent reasoning and decision-making processes.

\textbf{Reliability.}  The framework only outputs verified responses after the dynamically generated defense process has been executed and validated. Dedicated verification agents check the integrity of each reasoning step to ensure semantic correctness before producing the final output.

\textbf{Autonomy.} Each sub-agent autonomously schedules and executes defense tasks through an event-driven mechanism. A global coordination agent reconstructs workflows based on observed anomalies or feedback signals---activating, pausing, or rerouting defense modules. This self-regulating architecture enables continuous and adaptive defense.

\textbf{Training-free and Model-agnostic.}  \SystemName relies entirely on context-based learning and adaptive reasoning, requiring no fine-tuning or external training. It is model-agnostic and can be deployed on mainstream LLM-HARs to provide real-time defensive responses with minimal computational overhead.

Together, these principles enable \SystemName to function as a fully automated, self-defending reasoning layer---detecting, mitigating, and verifying prompt injection attacks while maintaining semantic accuracy and operational efficiency. In the following, we detail the design of our system.

\subsection{\SystemName Workflow}
\label{subsec:architecture}
Given an input IMU sequence $x$ and a potentially tampered prompt $\tilde{p}$ (constructed as described in Section~\ref{sec:attack-framework}), the system transforms $(x,\tilde{p})$ into a defensive prompt $\hat{p}$ and a secure prediction result $\hat{y}$ through six tightly coupled components:

\textbf{Input Sanitization}, \textbf{Consistency Verification}, \textbf{Robust Reasoning}, the \textbf{Memory Hub}, the \textbf{Planning Agent}, and the \textbf{Executor Agent}. We reuse the attack notation: $d = g(x)$ denotes the signal-to-text description of the IMU signal, $s_t$ and $c_t$ denote the benign instruction and context, $p = s_t \oplus d \oplus c_t$ is the benign prompt, and $f(\cdot)$ is the backend LLM.

\paragraph{(1) Input Sanitization.}
The Input Sanitization module is the first stage of the workflow and forwards raw inputs to the sanitization operators in the Memory Hub.

Given the raw pair $(x,\tilde{p})$, the module constructs the interaction tuple $\mathcal{I}=\{\mathcal{G},\mathcal{S},\mathcal{R},\mathcal{U},\mathcal{Y}_b\}$, where $\mathcal{G}$ is the current defense goal, $\mathcal{S}$ summarizes runtime state, $\mathcal{R}$ is a guard request encoding sanitization strictness, $\mathcal{U}$ is the observed prompt ($p$ or $\tilde{p}$), and $\mathcal{Y}_b=f(p)$ is the benign LLM output when available. If only $\tilde{p}$ is observed, a benign prototype $p$ is approximated from the Memory Hub to initialize $\mathcal{R}$.

Each component of $\mathcal{I}$ is embedded through the shared encoder $\Phi(\cdot)$ as $e_\xi=\Phi(\xi)\in\mathbb{R}^d$, enabling uniform comparison across signals, prompts, and state tokens~\cite{reimers2019sentencebert}. A lightweight deviation check $\mathrm{sim}(e_{\tilde{p}},e_{\text{benign}})<\tau_{\text{san}}$ with $\tau_{\text{san}}=0.75$ is performed before invoking the Hub, following standard similarity-based drift detection~\cite{garg2020drift}. Upon violation of the threshold, the module requests high-sensitivity sanitization.

Sanitization itself occurs inside the Memory Hub via $(\hat{x},\hat{p},\eta_{\text{san}})=\textsc{HubSanitize}(x,\tilde{p},\mathcal{I})$, where $\hat{x}$ and $\hat{p}$ are the sanitized IMU sequence and prompt, and $\eta_{\text{san}}$ contains sanitization metadata. Internally, \textsc{HubSanitize} applies MAD-based normalization with threshold $\tau_{\mathrm{MAD}}=2.5$~\cite{leys2013mad} and lexical filtering with graylist coefficient $\alpha=0.4$~\cite{raffel2020t5}.

From a workflow perspective, Input Sanitization simply prepares $\mathcal{I}$, calls the Hub, and forwards $(\hat{x},\hat{p},\eta_{\text{san}})$ to the next stage.

\paragraph{(2) Consistency Verification.}
The Consistency Verification module checks whether the purified inputs $\hat{x}$ and $\hat{p}$ remain consistent with the context in both semantic and temporal dimensions, while detecting cross-modal drift to identify residual adversarial effects. Similar to the Input Sanitization module, this component does not execute verification logic locally but instead delegates all operations to consistency operators provided by the memory hub.

Let $F_{\text{sens}}$ denote features extracted from the sanitized IMU signal $\hat{x}$ (e.g., statistical descriptors or learned embeddings), and let $C_{\text{ctx}}$ denote contextual attributes (environment, user state, interaction phase) derived from $\mathcal{S}$ and $\hat{p}$. Consistency Verification requests the hub to compute semantic and temporal consistency scores $(\gamma_{\text{sem}}, \gamma_{\text{temp}})=\textsc{HubCheckConsistency}(F_{\text{sens}}, C_{\text{ctx}}, \hat{x}, \hat{p})$, and when the system observes only a corrupted prompt $\tilde{p}$, it reconstructs an approximate benign prompt $p$ using prototype clean prompts retrieved from the Memory Hub, following prototype-based reconstruction strategies used in prior representation-learning systems~\cite{snell2017prototypical}. This reconstructed $p$ refines $C_{\text{ctx}}$ and informs the initialization of the guard request $\mathcal{R}$ based on the defense goal $\mathcal{G}$ and runtime state $\mathcal{S}$, ensuring that consistency checks remain reliable even under complete prompt corruption.

The semantic consistency indicator $\gamma_{\text{sem}}$ is produced by the hub’s conflict-scoring and rule-gating mechanism using the semantic mismatch threshold $\tau_c=0.35$, such that a semantic conflict is raised whenever $\Phi(F_{\text{sens}}, C_{\text{ctx}})>\tau_c$, consistent with semantic-gating heuristics~\cite{bao2022gptscore}. The temporal indicator $\gamma_{\text{temp}}$ is computed from the FastDTW alignment distance $d$ between the sanitized IMU trajectory and auxiliary contextual signals, mapped to the normalized cross-modal score~\cite{salvador2007fastdtw}.

If either the semantic or temporal threshold is violated, the module attaches a warning flag to the current interaction and updates the trust configuration propagated to the planning agent.

\paragraph{(3) Robust Reasoning.}
The Robust Reasoning Module controls how queries are made to the backend large language model $f(\cdot)$ and imposes constraints on its reasoning process under potential attacks. As the reasoning layer within a broader workflow encompassing planning agents, execution agents, and memory hubs, this module determines when to activate Robust Reasoning modes—such as chain-of-reasoning templates~\cite{wei2022cot}, multi-path voting via self-consistency~\cite{wang2023selfconsistency}, and veto mechanisms inspired by debate-style corrective inference~\cite{du2023improving,liang2022redteaming}—and routes reasoning through underlying Robust Reasoning operators. Robust reasoning is triggered when upstream inconsistency indicators exceed preset thresholds, specifically when the semantic conflict score satisfies $\gamma_{\text{sem}} > \tau_{\text{sem}}$ with $\tau_{\text{sem}} = 0.6$, or when the temporal consistency score falls below $\gamma_{\text{temp}} < \tau_{\text{temp}}$ with $\tau_{\text{temp}} = 0.6$.

Given the defended prompt $\hat{p}$ from the previous stages and the defense goal $\mathcal{G}$, Robust Reasoning prepares an inference configuration $\Theta_{\text{rob}}=(\mathcal{G},\mathbf{w}_{\text{mod}},\gamma_{\text{sem}},\gamma_{\text{temp}})$ and calls $(\hat{p},\hat{y},\eta_{\text{rob}})=\textsc{HubRobustInfer}(\hat{p},\Theta_{\text{rob}})$, where $\hat{y}$ denotes the final defended output of $f(\cdot)$, and $\eta_{\text{rob}}$ contains the reasoning trace, intermediate scores, and veto signals generated by the Memory Hub.

Internally, \textsc{HubRobustInfer} enforces a fixed chain-of-reasoning framework~\cite{wei2022cot}, concurrently executes multiple protected inference paths aggregated through self-consistency voting~\cite{wang2023selfconsistency}, and activates veto-based backtracking mechanisms~\cite{du2023improving,liang2022redteaming} when necessary. From a workflow perspective, this module optionally invokes Robust Reasoning when $\gamma_{\text{sem}}>0.6$ or $\gamma_{\text{temp}}<0.6$, passes the defended prompt and configuration to the Hub, and returns the final result for $(\hat{p},\hat{y})$ together with Robust Reasoning metadata for writing back to the Memory Hub.

\paragraph{(4) Memory Hub.}
The Memory Hub is a core component that implements all specific defense methods and maintains the defense condition knowledge base. It has two tightly coupled roles: a \textbf{Defense Function Library}, which contains a set of operators that have undergone purification, verification, and Robust Reasoning, and a \textbf{Knowledge Base}, which stores a repository of historical trajectories with cleaned and annotated defense results.

We denote the raw memory state as $\mathcal{M}_{\text{raw}}=\big\{(\mathcal{U}_t,\mathcal{Y}_t,\mathcal{P}_t,\mathcal{C}_t,\mathcal{R}_t)\big\}_{t=1}^T$, where $\mathcal{U}_t$ is the $t$-th observed prompt (benign or compromised), $\mathcal{Y}_t$ is the corresponding LLM response, $\mathcal{P}_t$ is the defense plan executed at time $t$, $\mathcal{C}_t$ contains intermediate corrective content, and $\mathcal{R}_t$ is the final repaired output verified as successful. Before any trajectory is made available to other modules, the Hub applies its internal defense operators to produce a protected memory state $\mathcal{M}'=\mathcal{D}(\mathcal{M}_{\text{raw}})$, where $\mathcal{D}(\cdot)$ is realized as a set of concrete methods described below. All retrievals and similarity computations are performed over $\mathcal{M}'$, not over $\mathcal{M}_{\text{raw}}$.

\textbf{(4.1) Input Sanitization operators inside the Memory Hub.}
For sensor-side sanitization, the Hub employs multiple normalization pipelines. Let $z_{ij}$ denote the $j$th normalized version of the $i$th configuration; the fused sanitized representation is computed as $\hat{z}_i=\operatorname{median}\!\left(\sum_j w_{ij}z_{ij}\right)$ with weights $w_{ij}\propto 1/\operatorname{MAD}(z_{ij})$ following the robust MAD criterion~\cite{leys2013mad}. Normalization channels exhibiting lower variability receive higher weights, which suppresses adversarially amplified artifacts while preserving legitimate motion patterns. For textual sanitization, the Hub uses a lexical filtering module based on the Aho--Corasick automaton~\cite{aho1975efficient}. For each lexical unit $w$, a trust score is assigned by $R(w)=0$ if $w\in\text{blacklist}$, $R(w)=\alpha$ if $w\in\text{greylist}$, and $R(w)=1$ otherwise, with $0\le\alpha<1$. Based on $R(w)$, the Hub removes or downweights suspicious tokens and outputs a sanitized prompt $p'$ together with a confidence profile. This $p'$ is exposed externally as $\hat{p}$ in the workflow.

\textbf{(4.2) Consistency verification operators inside the Memory Hub.}
To evaluate semantic consistency, the hub employs a rule-gated mismatch function $\Phi(F_{\text{sens}},C_{\text{ctx}})$ by comparing the perceived feature $F_{\text{sens}}$ with the contextual attribute $C_{\text{ctx}}$, and a semantic conflict is defined as $\text{Conflict}(F_{\text{sens}},C_{\text{ctx}})=1$ when $\Phi(F_{\text{sens}},C_{\text{ctx}})>\tau_c$ and $0$ otherwise, where $\tau_c$ is the adaptive threshold and semantic embedding alignment uses a Sentence-BERT–style encoder~\cite{reimers2019sentencebert}. Time-domain validation is achieved via FastDTW~\cite{salvador2007fastdtw}, where the alignment distance $d=\mathrm{DTW}(z_{\text{IMU}},s_{\text{aux}})$ is mapped to the normalized score $S_{\mathrm{cm}}=1/(1+d/T)$ using the cleaned IMU trajectory $z_{\text{IMU}}$, auxiliary signal $s_{\text{aux}}$, and window length $T$, and a temporal conflict is flagged when $S_{\mathrm{cm}}<0.6$. The Memory Hub subsequently applies weighting to inconsistent modalities and outputs semantic metrics $(\gamma_{\text{sem}})$ and temporal metrics $(\gamma_{\text{temp}})$ to upstream modules.

\textbf{(4.3) Robust Reasoning operators inside the Memory Hub.}
Memory Hub employs a structured reasoning framework defined as $R_{\text{CoT}}=\langle \text{statistics},\text{ prior checks},\text{ evidence reasoning},\text{ conclusions}\rangle$, inspired by chain-of-thought prompting~\cite{wei2022cot} and self-consistency reasoning~\cite{wang2023selfconsistency}. Multiple protected reasoning paths run in parallel: if $S_i$ is the score of the $i$-th chain, the aggregated score is $S_{\text{final}}=\operatorname{median}(S_1,S_2,\dots,S_n)$, and each chain outputs a veto signal $\text{veto}_i$ whose global form $V=\bigvee_{i=1}^n\text{veto}_i$ triggers backtracking and re-evaluation when $V=1$. All inference trajectories and scores are stored in $\mathcal{M}'$.

\textbf{(4.4) Retrieval and reuse.}
For the purified prompt $\hat{p}$, the Memory Hub embeds it as $e_u=\Phi(\hat{p})$ and retrieves memory entries according to $\mathcal{R}=\mathrm{TopK}(\mathrm{sim}(e_u,e_j))$, prioritizing examples with successful defense records following prototype-style retrieval principles~\cite{snell2017prototypical}.

\paragraph{(5) Planning Agent.}
The planner receives $(\hat{x}, \hat{p})$, validation metrics $(\gamma_{\text{sem}}, \gamma_{\text{temp}})$ and the retrieved memory set $\mathcal{R}$, determining which defense operators to invoke and their execution order. Although the planner does not directly perform language model inference, it relies on the semantic embedding space provided by the underlying Gemma-2-9B model~\cite{gemma2024google}, whose encoder serves as the shared feature mapping $\Phi(\cdot)$ for all text representations within the system.

The sanitized prompt is embedded as $e_u=\Phi(\hat{p})$ and compared against the benign prototype set $\mathcal{E}_b$ of the hub using the cosine similarity $\mathrm{sim}(e_u,\mathcal{E}_b)=\max_j \frac{e_u\cdot e_j}{\|e_u\|\cdot\|e_j\|}$. If this similarity score falls below a detection threshold or if the Hub flags inconsistencies through $(\gamma_{\text{sem}},\gamma_{\text{temp}})$, the Planner classifies the interaction as suspicious and constructs a defense plan $\mathcal{P}=\{\mathcal{T},\mathcal{D}_{\text{seq}},\mathcal{M}_{\text{cfg}},\mathcal{O}\}$, where $\mathcal{T}$ specifies the inferred threat type and $\mathcal{D}_{\text{seq}}$ contains symbolic defense steps (e.g., MADNormalize,
LexicalFilter, Canonicalize, CoTRobust), each corresponding to concrete operators implemented inside the Memory Hub.

The finalized plan $\mathcal{P}$ is subsequently forwarded to the execution agent, which translates the plan into specific calls to the hub, ultimately generating robust inference requests through the Gemma-2-9B model~\cite{gemma2024google}.

\paragraph{(6) Executor Agent.}
The execution agent receives the defense plan $\mathcal{P}$ and its associated memory set $\mathcal{R}$, translating the high-level symbolic plan into concrete invocations of the memory hub and underlying large language model. In our implementation, all inference queries are similarly executed by the Gemma-2-9B model~\cite{gemma2024google}.

The executor constructs a concrete sequence of function calls $\mathcal{F}_D=[f_1,f_2,\dots,f_m]$ with $f_i\in\textsc{DefenseOps}$, where each $f_i$ is a callable defense primitive registered in the Memory Hub operation library.

Let $h^{(0)}$ denote the initial internal representation derived from $(\hat{x},\hat{p})$. The Executor iteratively applies the update rule $h^{(i)}=f_i(h^{(i-1)},\mathcal{R},\mathcal{G})$, which is internally executed as the remote call $h^{(i)}=\textsc{HubCall}(f_i,h^{(i-1)},\mathcal{R},\mathcal{G})$. During execution, the executor monitors the semantic similarity between intermediate states and benign prototypes stored in the hub; when the condition $\mathrm{sim}(h^{(i)},h_{\text{base}})<\tau_{\text{exec}}$ (with $\tau_{\text{exec}}=0.85$) is satisfied, the executor upgrades to robust inference mode and invokes \textsc{HubRobustInfer}. This mechanism routes requests to Gemma-2-9B~\cite{gemma2024google} through chain-of-thought templates, multi-path voting, and veto protection.

Finally, the complete defense trajectory: The plan $\mathcal{P}$, call sequence $\mathcal{C}_D$, intermediate representations $\{h^{(i)}\}$, and final output $(\hat{p},\hat{y})$ are written back to the memory hub as a new entry in $\mathcal{M}_{\text{raw}}$, thereby closing the detection-purification-verification-inference loop.

As shown in Figure~\ref{fig:system_overview}(b), \SystemName employs a \textbf{multi-stage adaptive architecture} that integrates reasoning, memory retrieval, and automated defense execution into a unified detect-remedy-verify loop. The system comprises two cooperating AI agents—a planner agent and an executor agent—communicating through a shared memory hub. Together, these components form a closed-loop reasoning-defense cycle capable of autonomously detecting, mitigating, and verifying prompt injection threats.

\section{Evaluation}
\label{sec:evaluation}
\subsection{Goals, Metrics, and Methodologies} 
The goal of the evaluation is to answer:
\begin{itemize}
  \item \textbf{RQ1 (Defense effectiveness):} How accurate is \SystemName in prompt attack detection and correction on different LLMs and datasets?
   \item \textbf{RQ2 (Ablation study):} What is the specific contribution of each defensive component, namely, input sanitizer, consistency verifier, and robust reasoner?
  \item \textbf{RQ3 (Attack severity):} Among the three attack paths in Sec.~\ref{subsubsec:attackmethods}, which attack poses the highest level of severity?
  \item \textbf{RQ4 (Comparison with baselines):} Does \SystemName outperform state-of-the-art defenses against prompt injection attacks?
   \item \textbf{RQ5 (Latency):} What is the runtime overhead of \SystemName? Will it cause significant delay when deployed with popular LLM-HARs?

\end{itemize}

\textbf{Target Models.} 
We conduct extensive evaluation on five state-of-the-art LLM-HARs:

\begin{itemize}
   \item \textbf{IMUGPT-2.0}~\cite{leng2024imugpt2}.
    IMUGPT-2.0 is a prompt-based IMU activity recognizer that encodes IMU features into compact text prompts for classification by LLMs.
    
    \item \textbf{MotionGPT}~\cite{jiang2023motiongpt}. 
    MotionGPT is a unified action-language generation model capable of learning bidirectional mappings between human action sequences and natural language descriptions.

    \item \textbf{HARGPT}~\cite{yang2024hargpt}.
    HARGPT achieves zero-shot IMU activity inference through purely natural language prompts, relying on LLM reasoning rather than supervised classifiers.


    \item \textbf{LLaSA}~\cite{imran2024llasa}.
    LLaSA achieves sensor-driven narrative generation and perceptual question-answering by integrating IMU features with natural language reasoning.

    \item \textbf{ContextGPT}~\cite{arrotta2024contextgpt}.
    ContextGPT achieves context-aware activity inference by integrating contextual knowledge (i.e., location, daily activities, and environment) with IMU summaries.
\end{itemize}

We use the following public HAR datasets to train the above target models (not \SystemName which is training-free):
\begin{itemize}
    \item \textbf{USC-HAD}~\cite{usc-had}.
    This wearable motion dataset contains 12 daily activities, sourced from human-worn accelerometer and gyroscope sensors.
    We select USC-HAD as a representative IMU dataset to evaluate the impact of prompt injection on typical short-duration human activities.
    \item \textbf{UCI HAR}~\cite{anguita2013public}.
    A smartphone-based HAR dataset featuring accelerometer and gyroscope recordings during six common activities.
    Its sensor configuration and mounting positions differ from USC-HAD, enabling cross-heterogeneous IMU source evaluation.
    \item \textbf{PAMAP2}~\cite{pamap2}.
    This multi-sensor HAR dataset encompasses 18 diverse physical activities, collected from multiple IMU locations.
    Its rich activity types and multi-location signals enable testing of \SystemName under more complex and refined motion patterns.
\end{itemize}

Since the output of these target models are different, we feed their output into a LLM-based HAR classifier for evaluation consistency. Specifically, we design four LLM-based HAR classifiers whose outputs are activity labels, such as running, walking, and standing. These LLMs include ChatGPT-4o, Gemini, LLaMA-2, and DeepSeek-V3. By standardizing the LLM outputs, we use the following metrics commonly used in previous prompt attack studies~\cite{greshake2023more, zou2023universal, guo2024prompt}.

\begin{itemize}
    \item \textbf{Detection Accuracy (DA).}
    The DA metric evaluates \SystemName's ability to correctly identify adversarial inputs before they propagate to downstream HAR pipelines. 

    \item \textbf{Attack Success Rate (ASR).}
    ASR quantifies how many originally correct predictions become incorrect after an attack. 
   
    \item \textbf{Recovery Rate (RR).}
    RR measures how effectively \SystemName recovers errors introduced by the attack. 

    \item \textbf{Semantic Consistency (SC).}
    SC reflects the proportion of defense outputs that remain semantically aligned with the original model outputs. 
    
    \item \textbf{Harm Score (HS).}
    HS measures the potential risk posed by a model's erroneous predictions under adversarial prompts. This metric quantifies the severity of harm caused by misclassification relative to the underlying physical activity. Each prediction is assigned a discrete hazard level $s_i \in \{1,2,3,4,5\}$ based on the safety-critical distance between the predicted activity and the true label, with higher values indicating more severe consequences.
\end{itemize}


\textbf{Baselines.} We compare \SystemName with:
\begin{itemize}
    \item \textbf{Text-only defenses.} SafeDecoding~\cite{zeng2024safedecoding} is a representative method that detect malicious intent through real-time classifiers and mitigate risks by decoding and rewriting.

    \item \textbf{Classical HAR defenses.}
    Traditional IMU defense mechanisms, such as denoising and adversarial training against FGSM/PGD~\cite{fgsm2015,pgd2018}, are effective against perturbations of raw signals but do not address language models.

    \item \textbf{Multimodal detection-only defenses.}
    These methods rely on cross-modal consistency checks or anomaly detection, such as~\cite{cliprobust2022,multimodal_adversarial2020}, to flag mismatched sensor-text pairs.
\end{itemize}

\textbf{Experimental Setup.}
All evaluations are conducted using default configurations of target LLM-HAR models. To minimize randomness, each experiment is repeated five times, with average results reported. All the evaluation are ran on a workstation equipped with three NVIDIA RTX 3090 GPUs (24GB memory each), ensuring sufficient computational resources for consistent evaluation across all baselines and attack scenarios.

\subsection{Defense Effectiveness (RQ1)}
Table~\ref{tab:rq1_avg} summarizes the average performance of \SystemName across different models. Overall, \SystemName achieves an average DA of 85\% and reduces ASR by 30\% on average, demonstrating robust defense capabilities. Additionally, \SystemName attains an average semantic consistency score of 75\%, indicating its ability to effectively correct injected tampering while preserving the original intent of the prompt. Among the evaluated models, IMUGPT-2.0 exhibits the highest vulnerability to adversarial prompts. In contrast, MotionGPT demonstrates the strongest resistance to prompt injection attacks: its ASR drops by approximately 29\% when \SystemName is enabled. In LLM-based HAR classifiers, GPT-4o and Gemini consistently demonstrate the strongest robustness against adversarial inputs, while DeepSeek and LLaMA2 exhibit greater vulnerability. Nevertheless, \SystemName delivers sustained, significant defensive improvements across all combinations of target models, datasets, and down-stream LLM-based HAR classifiers, substantially enhancing the system's stability against diverse prompt injection attacks.

\begin{table}[t]
\centering
\caption{\textbf{Overall performance of \SystemName on five target models.}
Metrics include Detection Accuracy (DA), Semantic Consistency (SC), and Attack Success Rate
(ASR, without \SystemName~$\rightarrow$~with \SystemName).}
\label{tab:rq1_avg}
\tiny
\setlength{\tabcolsep}{4pt}
\renewcommand{\arraystretch}{0.95}

\begin{tabularx}{\linewidth}{l l c c c}
\toprule
\textbf{Model} & \textbf{HAR classifier} &
\textbf{DA (\%)} &
\textbf{SC (\%)} &
\textbf{ASR (w/o $\rightarrow$ w/ \SystemName)} \\
\midrule

\multirow{4}{*}{IMUGPT-2.0}
 & GPT-4o   & 82.3 & 76.0 & 62.2\% $\rightarrow$ 33.4\% \\
 & Gemini   & 81.2 & 74.2 & 63.5\% $\rightarrow$ 37.4\% \\
 & DeepSeek & 82.3 & 74.5 & 59.7\% $\rightarrow$ 34.6\% \\
 & LLaMA2   & 78.5 & 69.2 & 59.0\% $\rightarrow$ 41.5\% \\
\midrule

\multirow{4}{*}{MotionGPT}
 & GPT-4o   & 84.2 & 78.5 & 62.2\% $\rightarrow$ 30.0\% \\
 & Gemini   & 83.2 & 77.0 & 63.5\% $\rightarrow$ 33.0\% \\
 & DeepSeek & 83.8 & 77.0 & 59.7\% $\rightarrow$ 31.4\% \\
 & LLaMA2   & 80.3 & 72.2 & 59.0\% $\rightarrow$ 35.2\% \\
\midrule

\multirow{4}{*}{LLaSA}
 & GPT-4o   & \textbf{90.5} & 78.0 & 62.2\% $\rightarrow$ 27.8\% \\
 & Gemini   & 86.5 & 76.0 & 63.5\% $\rightarrow$ 29.9\% \\
 & DeepSeek & 87.0 & 76.0 & 59.7\% $\rightarrow$ 30.0\% \\
 & LLaMA2   & 83.3 & 72.2 & 59.0\% $\rightarrow$ 34.1\% \\
\midrule

\multirow{4}{*}{HAR-GPT}
 & GPT-4o   & 86.2 & 77.0 & 62.2\% $\rightarrow$ 29.0\% \\
 & Gemini   & 84.2 & 75.0 & 63.5\% $\rightarrow$ 30.7\% \\
 & DeepSeek & 84.9 & 75.0 & 59.7\% $\rightarrow$ 31.8\% \\
 & LLaMA2   & 81.5 & 71.5 & 59.0\% $\rightarrow$ 35.2\% \\
\midrule

\multirow{4}{*}{ContextGPT}
 & GPT-4o   & \textbf{93.0} & 79.0 & 62.2\% $\rightarrow$ 27.2\% \\
 & Gemini   & 89.0 & 77.0 & 63.5\% $\rightarrow$ 29.0\% \\
 & DeepSeek & 89.8 & 77.0 & 59.7\% $\rightarrow$ 29.3\% \\
 & LLaMA2   & 84.2 & 73.2 & 59.0\% $\rightarrow$ 33.1\% \\
\bottomrule
\end{tabularx}
\end{table}

\subsection{Ablation Study (RQ2)}
\label{sec:ablation-rq2}
To quantify the contributions of the three defense components—input sanitizer, consistency verifier, and robust reasoner—in the \SystemName system, we conduct an ablation study by sequentially disabling each module and evaluating the performance of all variants on five target models. Figure~\ref{fig:bar-ablation} illustrates the DA under different attack types. Collectively, these results highlight the complementary nature of the three components. The full version of \SystemName demonstrates the strongest and most stable robustness, consistently maintaining high detection accuracy across all five LLM-HAR target models. In contrast, all stripped-down versions exhibit significant performance degradation, with vulnerabilities exposed directly corresponding to the removed functionalities. Removing the Input Sanitizer results in a moderate decrease in accuracy, lowering the DA to the 53--57\% range. This decline highlights the importance of early normalization in mitigating formatting errors or noisy instruction patterns. Disabling the Consistency Verifier further reduces robustness, dropping DA to 46--49\%. This decline highlights the critical role of cross-modal semantic alignment in preventing misinterpretations between sensor-generated motion descriptions and textual prompts. Finally, removing the robust reasoner causes performance to crash dramatically, with DA plummeting to just 7--8\%. This severe failure demonstrates that the robust reasoner is essential for maintaining task intent and reconstructing coherent prompts when confronted with structural perturbations, contextual contamination, or disruptions to higher-level reasoning.

\begin{figure}[t]
    \centering
    \includegraphics[width=0.89\linewidth]{fig_ablation_bar.pdf}
    \caption{Detection accuracy of different variants: full system, without Input Sanitizer, without Consistency Verifier, and without Robust Reasoner.}
    \label{fig:bar-ablation}
    \vspace{-0.2in}
\end{figure}

Among all fifteen attack types, the complete \SystemName model consistently outperforms all truncated variants. Their synergistic interactions stabilize lexical variability, intercept deeper semantic distortions, and reconstruct coherent prompts under structural manipulation, enabling \SystemName to exhibit the most robust and universally effective defensive behavior.

\subsection{Attack Severity (RQ3)}
\label{subsec:attackseverity}

In this experiment, we examine the safety-critical impact of the fifteen instantiated attacks under the unified operator $A(\cdot)$. 
Table~\ref{tab:hs_attacks} reports the averaged harm score (HS) under the original 1--5 scale, aggregated across all five LLM-HAR pipelines and three datasets. 

We observe a distinct hierarchy among the four attack categories. Text path attacks occupy the lower end of the severity spectrum, with HS values generally clustered between 2.6 and 2.8. This indicates that while linguistic perturbations can mislead LLMs, they rarely cause significant semantic shifts in underlying activity explanations. Prompt path attacks exhibit slightly higher severity, scoring approximately 2.9--3.1, as tampering with task intent or inserting conflicting contextual cues more directly disrupts the inference stage. The severity of signal path attacks exhibits a more pronounced escalation, with their HS values rising to approximately 3.5--3.6. Since these attacks directly manipulate the IMU data stream, they distort the physical dynamics of motion itself, thereby amplifying downstream transformation and inference errors. Hybrid attacks pose the most severe threat, with HS values reaching 3.8--4.1. By simultaneously perturbing both the kinetic properties at the sensor level and the semantic information at the prompt level, these attacks create a compounding effect, consistently generating the most dangerous prediction error patterns across all evaluated LLM-HAR attack pipelines.

\begin{table}[t]
\centering
\caption{Harm Score (HS, 1--5) for the fifteen adversarial attacks under the unified operator $A(\cdot)$. Higher HS indicates more severe safety-critical mispredictions.}
\label{tab:hs_attacks}
\scriptsize
\renewcommand{\arraystretch}{0.95}

\begin{tabularx}{\linewidth}{l X c c}
\toprule
\textbf{Category} & \textbf{Attack Name} & \textbf{HS (1--5)} & \textbf{Avg.} \\
\midrule

\multirow{2}{*}{Signal Attacks} 
    & Noise Injection & 3.5 & \multirow{2}{*}{3.55} \\
    & Drift Attack    & 3.6 & \\
\midrule

\multirow{2}{*}{Text Attacks}
    & Synonym Bias          & 2.6 & \multirow{2}{*}{2.70} \\
    & Adversarial Rewriting & 2.8 & \\
\midrule

\multirow{8}{*}{Prompt Attacks}
    & Prompt Concatenation & 2.9 & \multirow{8}{*}{3.00} \\
    & Task Injection       & 3.0 & \\
    & Role Confusion       & 3.0 & \\
    & CoT Interference     & 3.1 & \\
    & Multi-Task Blending  & 3.1 & \\
    & Context Pollution    & 3.0 & \\
    & Label Mixing         & 2.9 & \\
    & Unrelated Text Noise & 3.0 & \\
\midrule

\multirow{3}{*}{Hybrid Attacks}
    & Few-shot Poisoning & 3.8 & \multirow{3}{*}{3.97} \\
    & Semantic Drift     & 4.0 & \\
    & Hybrid Combo       & 4.1 & \\
\bottomrule
\end{tabularx}
\vspace{-0.2in}
\end{table}

\begin{figure}[t]
    \centering
    \includegraphics[width=0.84\linewidth]{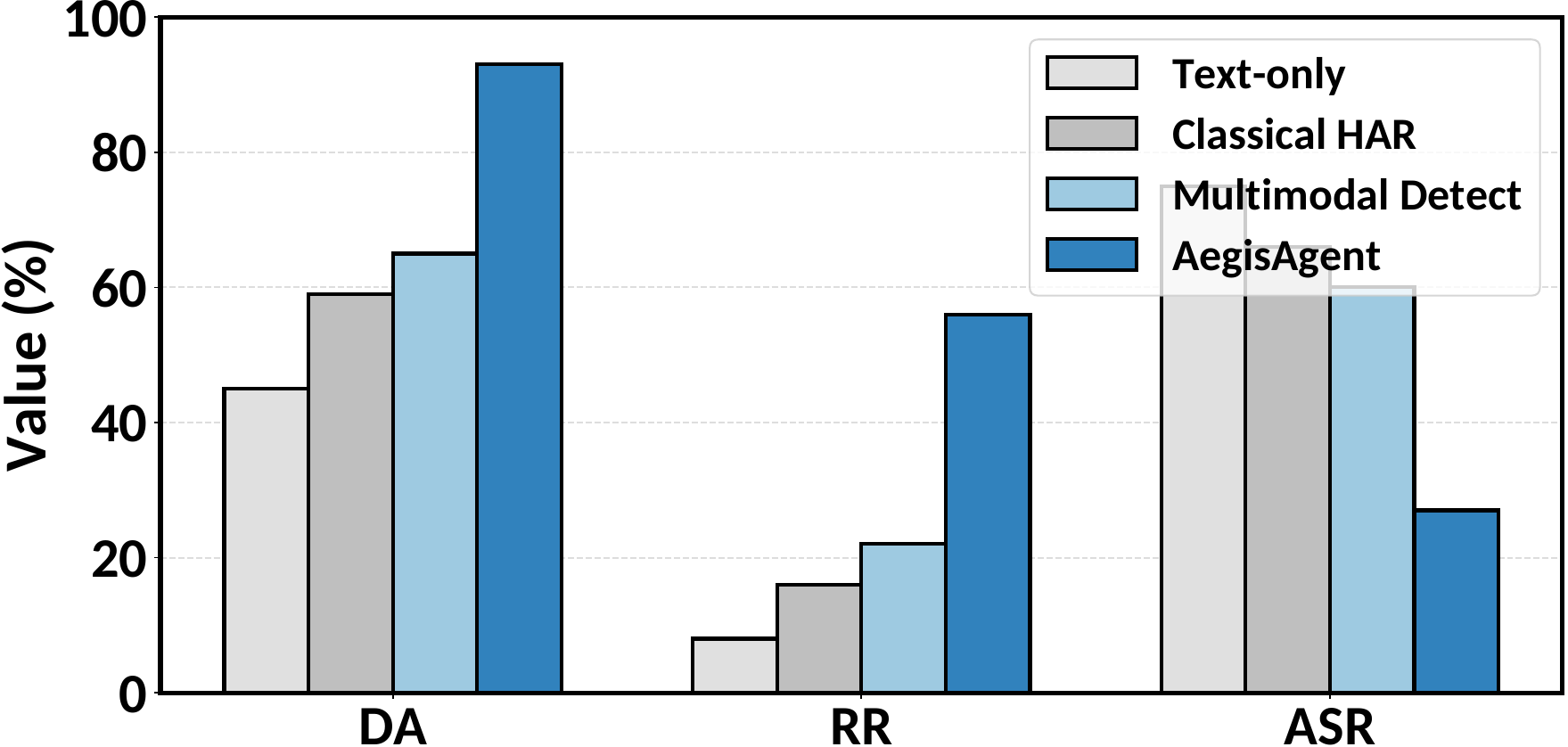}
    \vspace{-1pt}  
    \caption{
        Performance comparison between AegisAgent and three defense baselines: text-only defense, 
        classical HAR defense, and multimodal detection-only defense. Metrics include DA, RR, and ASR.
    }
    \label{fig:rq4_baseline}
\end{figure}

\subsection{Comparison with baselines (RQ4)}
\label{subsec:comparison}
We now compare \SystemName with state-of-the-art defenses against prompt attacks against LLMs. Figure~\ref{fig:rq4_baseline} summarizes the average values of DA, RR, and ASR across all target models. Text-only defenses demonstrate the weakest protection, recovering only partial damaged prompts and resulting in the success of most adversarial attacks, achieving only 44.1\% DA and 7.9\% RR, while leaving ASR as high as 74.3\%. Classical HAR defenses perform slightly better by mitigating low-level IMU perturbations, yet remains ineffective against semantic prompt tampering; their performance increases to only 58.7\% DA and 15.8\% RR, with ASR still at 66.1\%. Multimodal detection-only defenses further improves DA and RR, reaching 64.9\% DA and 21.7\% RR, yet remains vulnerable to strong prompt injection since it detects but cannot repair semantic inconsistencies, maintaining an ASR of 60.4\%. In contrast, \SystemName demonstrates significantly stronger robustness across all metrics. It achieves higher detection accuracy (93.0\% DA), substantially enhanced recovery capability (56.3\% RR), and more effectively reduces ASR than all baselines, lowering it to only 27.2\%. Overall, \SystemName delivers the most consistent and comprehensive defense, demonstrating that combining input sanitization, cross-modal consistency verification, and robust reasoning is crucial for protecting LLM-based HAR systems against multi-path prompt injection attacks.

\subsection{Latency (RQ5)}
\label{sec:resource}
We evaluate the runtime overhead introduced by \SystemName. To quantify the practical integration cost of the proxy-based defense scheme, we measured the single-query latency for each module in the full pipeline. The reported data reflects the actual additional runtime overhead when \SystemName runs in parallel with the LLM-HAR system. Table~\ref{tab:breakdown} summarizes module latencies: Input Sanitization via normalization and Aho-Corasick matching takes 6.3 ms; MiniLM-based embedding extraction adds 12.7 ms; cross-modal Consistency Verification using FastDTW and $\Phi$ contributes 18.4 ms. Memory Hub retrieval takes 9.6 ms, while the Planning Agent and Executor Agent adds 11.2 ms and 12.0 ms, respectively. Robust Reasoning incurs an overhead of 8.4 ms. To sum up, \SystemName adds an average latency of 78.6 ms per query on a NVIDIA RTX 3090 GPU workstation, indicating real-time responsiveness when deployed in a server.

\begin{table}[t]
\centering
\caption{Running time measured on the GPU workstation.}
\label{tab:breakdown}
\scriptsize
\begin{tabularx}{0.98\linewidth}{l l >{\centering\arraybackslash}X}
\toprule
\textbf{Main Module} & \textbf{Submodule} & \textbf{Time (ms)} \\
\midrule

\textbf{Input Sanitization}      
& Sanitizer (Canon.\ + Aho--Corasick)     
& 6.3  \\
\midrule

\textbf{Consistency Verification} 
& Embedding + Consistency Checking 
& 31.1 \\
\midrule

\multirow{4}{*}{\textbf{Robust Reasoning}}
& Memory Hub Retrieval                     
& 9.6  \\
& Planning Agent                            
& 11.2 \\
& Executor Agent                            
& 12.0 \\
& Robust Reasoner (20\% activation)        
& 8.4  \\
\midrule

\textbf{Total}          
&                                      
& \textbf{78.6} \\
\bottomrule
\end{tabularx}
\end{table}

\section{Related Work}
\label{sec:relatedwork}
\textbf{LLM-based HAR.}
Recent studies have explored the potential of utilizing LLMs to interpret IMU signals for HAR. Examples include IMUGPT-2.0~\cite{leng2024imugpt2}, MotionGPT~\cite{jiang2023motiongpt}, LLaSA~\cite{imran2024llasa}, HAR-GPT~\cite{yang2024hargpt}, and ContextGPT~\cite{arrotta2024contextgpt}. Together, they demonstrate that LLMs can generate action descriptions, integrate multimodal representations, and achieve cross-domain generalization across sensor modalities. These approaches primarily focus on enhancing activity interpretation capabilities, multimodal alignment accuracy, or zero-shot generalization. However, existing research generally assumes a benign and uninterrupted sensor-to-text processing pipeline, overlooking the interactive dependency between sensor-generated descriptions and user-provided prompts.

\textbf{Prompt attacks.}
A growing body of research indicates that large language models can be manipulated through malicious prompts. Previous studies on jailbreak suffixes~\cite{zou2023universal}, adversarial prompts~\cite{qi2023visual}, semantic drift attacks~\cite{greshake2023more}, and multimodal jailbreak scenarios~\cite{li2024mmjailbreak} demonstrate that even state-of-the-art LLMs remain susceptible to prompt-induced behavioral shifts. In-depth analysis further reveals that subtle contextual cues or reasoning disruptions can significantly alter model outputs, exposing fundamental vulnerabilities in text-driven interfaces. Existing research also indicates that deceptive phrasing~\cite{perez2023ignore} or chain-of-thought steering~\cite{turpin2023language} may undermine prompt robustness. However, these studies overlook the dual-source nature of adversarial perturbations in LLM-HAR systems, where threats may originate simultaneously from user commands and sensor-generated action descriptions.

Unlike previous studies, our work focuses on the overlooked vulnerability of cross-modal prompt manipulation within LLM-HAR pipelines. Specifically, we propose an automated detection and remediation solution targeting hierarchical attacks on prompts that influence both user instructions and IMU descriptions. To our knowledge, this is the first end-to-end defense mechanism specifically designed for multimodal LLM-HAR systems.

\section{Conclusion}
\label{sec:conclusion}

This paper introduces \SystemName---an automated agent defense solution for LLM-HAR systems. It aims to protect LLM-HAR systems from prompt injection attacks. This solution detects and repairs adversarial prompts without human intervention. It mitigates security risks arising from erroneous predictions in real-world scenarios and demonstrates robustness against a wide range of adversarial perturbations. Research indicates that \SystemName's automated defense mechanism is crucial for ensuring the security of LLM-HAR systems.

\bibliographystyle{unsrt}
\bibliography{ref}

@misc{openai2023gpt4,
  title        = {GPT-4 Technical Report},
  author       = {{OpenAI}},
  year         = {2023},
  howpublished = {\url{https://openai.com/research/gpt-4}},
}

@misc{anthropic2023claude,
  title        = {Claude: Constitutional AI},
  author       = {{Anthropic}},
  year         = {2023},
  howpublished = {\url{https://www.anthropic.com}},
}

@misc{google2023gemini,
  title        = {Gemini: A Family of Highly Capable Multimodal Models},
  author       = {{Google DeepMind}},
  year         = {2023},
  howpublished = {\url{https://deepmind.google/technologies/gemini}},
}

@article{bubeck2023sparks,
  title={Sparks of Artificial General Intelligence: Early Experiments with GPT-4},
  author={Bubeck, S{\'e}bastien and Chandrasekaran, Varun and Eldan, Ronen and Gehrke, Johannes and Horvitz, Eric and Kamar, Ece and Lee, Peter and Levy, Ofer and Li, Yossi and Morris, Roy and Munteanu, Andrei and Nori, Harsha and Palangi, Hamid and Prakash, Sahitya and Sontag, David and Weimer, Stefan and Zhang, Yi},
  journal={arXiv preprint arXiv:2303.12712},
  year={2023}
}

@article{touvron2023llama,
  title={LLaMA: Open and Efficient Foundation Language Models},
  author={Touvron, Hugo and Lavril, Thibaut and Izacard, Gautier and Martinet, Louis and Lachaux, Marie-Anne and Lacroix, Timoth{\'e}e and Rozi{\`e}re, Baptiste and Goyal, Naman and Hambro, Eric and Azhar, Faisal and Rodriguez, Aurelien and Joulin, Armand and Grangier, David and Bojanowski, Piotr and Lample, Guillaume},
  journal={arXiv preprint arXiv:2302.13971},
  year={2023}
}

@article{raffel2020t5,
  title={Exploring the Limits of Transfer Learning with a Unified Text-to-Text Transformer},
  author={Raffel, Colin and Shazeer, Noam and Roberts, Adam and Lee, Katherine and Narang, Sharan and Matena, Michael and Zhou, Yanqi and Li, Wei and Liu, Peter J.},
  journal={Journal of Machine Learning Research},
  volume={21},
  number={140},
  pages={1--67},
  year={2020}
}

@article{leng2024imugpt,
  title={IMUGPT: Zero-shot IMU-based Human Activity Understanding with LLMs},
  author={Leng, Zikang and Li, Haotian and Zhang, Ming and Wang, Zihan and He, Haoran},
  journal={arXiv preprint arXiv:2401.01234},
  year={2024}
}

@article{leng2024imugpt2,
author = {Leng, Zikang and Bhattacharjee, Amitrajit and Rajasekhar, Hrudhai and Zhang, Lizhe and Bruda, Elizabeth and Kwon, Hyeokhyen and Pl\"{o}tz, Thomas},
title = {IMUGPT 2.0: Language-Based Cross Modality Transfer for Sensor-Based Human Activity Recognition},
year = {2024},
issue_date = {September 2024},
publisher = {Association for Computing Machinery},
address = {New York, NY, USA},
volume = {8},
number = {3},
month = sep,
articleno = {112},
numpages = {32},
}

@article{jiang2023motiongpt,
  title={MotionGPT: Human Motion as a Foreign Language},
  author={Jiang, Yuxiao and Li, Wen and Liu, Tianyu and Hong, Hao and Komura, Taku and Liu, Ziwei},
  journal={arXiv preprint arXiv:2306.08993},
  year={2023}
}

@article{yang2024hargpt,
  title={HAR-GPT: Harnessing LLMs for IMU-based Human Activity Recognition},
  author={Yang, Zhiwei and Chen, Jie and Zhao, Rui and Xu, Cheng and Liu, Jing},
  journal={arXiv preprint arXiv:2402.02354},
  year={2024}
}

@article{imran2024llasa,
  title={LLASA: Large Language and Activity Sensing Agents},
  author={Imran, Ali and Xu, Yichuan and Kim, Dohyun and Chen, Longqi and Zhang, Hao and Gao, Yu},
  journal={arXiv preprint arXiv:2403.04567},
  year={2024}
}

@article{arrotta2024contextgpt,
  title={ContextGPT: Context-aware LLM for Human Activity},
  author={Arrotta, Gabriele and Saponara, Sergio and Marinoni, Maurizio},
  journal={arXiv preprint arXiv:2403.07777},
  year={2024}
}

@inproceedings{kenton2019bert,
  title={BERT: Pre-training of Deep Bidirectional Transformers for Language Understanding},
  author={Devlin, Jacob and Chang, Ming-Wei and Lee, Kenton and Toutanova, Kristina},
  booktitle={Proceedings of the 2019 Conference of the North American Chapter of the Association for Computational Linguistics (NAACL)},
  pages={4171--4186},
  year={2019}
}

@misc{radford2019language,
  title={Language Models are Unsupervised Multitask Learners},
  author={Radford, Alec and Wu, Jeffrey and Child, Rewon and Luan, David and Amodei, Dario and Sutskever, Ilya},
  year={2019},
  note={OpenAI Technical Report}
}

@inproceedings{zellers2021pig,
  title={PIGLeT: Language Grounding Through Heterogeneous Demonstrations},
  author={Zellers, Rowan and Holtzman, Ari and Peters, Matthew E. and Narang, Sharan and Eibling, Robert and Bordes, Antoine and Kiela, Douwe and Rush, Alexander},
  booktitle={Proceedings of the IEEE/CVF International Conference on Computer Vision (ICCV)},
  pages={1855--1865},
  year={2021}
}

@article{perez2022ignore,
  title={Ignore Previous Prompt: Attack Techniques for Prompt-based LLMs},
  author={Perez, Ethan and Rando, Juliana and Kiela, Douwe},
  journal={arXiv preprint arXiv:2211.09527},
  year={2022}
}

@article{greshake2023more,
  title={More Than You’ve Asked For: A Comprehensive Analysis of Jailbreak Attacks Against Large Language Models},
  author={Greshake, Kai and W{\"u}chner, Tobias and Wolf, Thomas},
  journal={arXiv preprint arXiv:2309.10253},
  year={2023}
}

@article{zhuo2023tptu,
  title={TPTU: Universal and Transferable Prompt Injection Attacks},
  author={Zhuo, Terry and Shen, Yao and Lin, Zeqi and Zhang, Tianyi and Cao, Shiqi and Sun, Zhenhua},
  journal={arXiv preprint arXiv:2305.15524},
  year={2023}
}

@inproceedings{papernot2016towards,
  title={Towards the Science of Security and Privacy in Machine Learning},
  author={Papernot, Nicolas and McDaniel, Patrick and Wu, Xi and Jha, Somesh and Swami, Ananthram and Beyah, Raheem},
  booktitle={2016 IEEE European Symposium on Security and Privacy (EuroS\&P)},
  pages={399--414},
  year={2016}
}

@inproceedings{zhang2020dolphin,
  title={DolphinAttack: Inaudible Voice Commands},
  author={Zhang, Guoming and Yan, Chen and Ji, Xiaoyu and Zhang, Tianchen and Zhang, Wenyuan and Xu, Wenyuan},
  booktitle={Proceedings of the ACM SIGSAC Conference on Computer and Communications Security (CCS)},
  pages={1032--1045},
  year={2017}
}

@inproceedings{goodfellow2015explaining,
  title={Explaining and Harnessing Adversarial Examples},
  author={Goodfellow, Ian and Shlens, Jonathon and Szegedy, Christian},
  booktitle={International Conference on Learning Representations (ICLR)},
  year={2015}
}

@article{lu2023sensor,
  title={Sensor-Based Adversarial Attacks on IMU Activity Recognition},
  author={Lu, Yuan and Yuan, Ming and Chen, Qi and Zhang, Xiangyu},
  journal={IEEE Transactions on Mobile Computing},
  year={2023}
}

@article{baltruvsaitis2018multimodal,
  title={Multimodal Machine Learning: A Survey},
  author={Baltrušaitis, Tadas and Ahuja, Chaitanya and Morency, Louis-Philippe},
  journal={IEEE Transactions on Pattern Analysis and Machine Intelligence},
  volume={41},
  number={2},
  pages={423--443},
  year={2018}
}

@inproceedings{ngiam2011multimodal,
  title={Multimodal Deep Learning},
  author={Ngiam, Jiquan and Khosla, Aditya and Kim, Mingyu and Nam, Juhan and Lee, Honglak and Ng, Andrew Y.},
  booktitle={Proceedings of the 28th International Conference on Machine Learning (ICML)},
  pages={689--696},
  year={2011}
}

@article{zou2023universal,
  title={Universal and Transferable Adversarial Attacks on Aligned Language Models},
  author={Zou, Andy and Wang, Zhen and Carlini, Nicholas and Nasr, Milad and Kolter, Zico and Fredrikson, Matt and Papernot, Nicolas},
  journal={arXiv preprint arXiv:2307.15043},
  year={2023}
}

@inproceedings{anguita2013public,
  title={A Public Domain Dataset for Human Activity Recognition Using Smartphones},
  author={Anguita, Davide and Ghio, Alessandro and Oneto, Luca and Parra, Xavier and Reyes-Ortiz, Jorge Luis},
  booktitle={21st European Symposium on Artificial Neural Networks, Computational Intelligence and Machine Learning (ESANN)},
  year={2013}
}

@inproceedings{usc-had,
  title={USC-HAD: A Daily Activity Dataset for Ubiquitous Activity Recognition Using Wearable Sensors},
  author={Zhang, Min and Sawchuk, Alexander A.},
  booktitle={Proceedings of the ACM Conference on Ubiquitous Computing (UbiComp)},
  pages={1036--1043},
  year={2012}
}

@inproceedings{pamap2,
  title={Physical Activity Monitoring for Aging People (PAMAP2) Dataset},
  author={Reiss, Attila and Stricker, Didier},
  booktitle={Proceedings of the 4th International Conference on Pervasive Computing Technologies for Healthcare (PervasiveHealth)},
  year={2012}
}

@article{gupta2024complexprompt,
  title={Complex Prompt Attacks on LLMs},
  author={Gupta, Amanpreet and Rando, Juliana and Kim, Dongyeop and Kiela, Douwe},
  journal={arXiv preprint arXiv:2402.01822},
  year={2024}
}

@article{yan2015sensor,
  title={Sensor-based Attacks on Mobile Devices: A Survey},
  author={Yan, Qiang and Yu, Qiang and Li, Fei and Jin, Yuqing},
  journal={IEEE Communications Surveys \& Tutorials},
  volume={17},
  number={3},
  pages={1439--1452},
  year={2015},
  publisher={IEEE}
}

@inproceedings{cao2021advimu,
  title={AdvIMU: Adversarial Attacks on IMU-based Activity Recognition Systems},
  author={Cao, Jie and He, Zheng and Chen, Xiaoyu and Li, Xiaohui and Wang, Wei},
  booktitle={Proceedings of the ACM Conference on Security and Privacy in Wireless and Mobile Networks (WiSec)},
  pages={106--116},
  year={2021}
}

@inproceedings{ma2019characterizing,
  title={Characterizing Sensor-based Attacks on Mobile Devices},
  author={Ma, Shiqi and Meng, Weizhi and Xiao, Han and Zhang, Xiangyu and Zhang, Feng},
  booktitle={Proceedings of the 2019 IEEE Symposium on Security and Privacy (S\&P)},
  pages={222--238},
  year={2019},
  publisher={IEEE}
}

@article{reimers2019sentencebert,
  title={Sentence-BERT: Sentence Embeddings using Siamese BERT-Networks},
  author={Reimers, Nils and Gurevych, Iryna},
  journal={Proceedings of the 2019 Conference on Empirical Methods in Natural Language Processing (EMNLP)},
  pages={3982--3992},
  year={2019},
  publisher={ACL}
}

@article{garg2020drift,
  title={Detecting Concept Drift in Text Data Streams},
  author={Garg, Siddhant and Agarwal, Surajit},
  journal={Proceedings of the 58th Annual Meeting of the Association for Computational Linguistics (ACL)},
  pages={1149--1159},
  year={2020},
  publisher={ACL}
}

@article{leys2013mad,
  title={Detecting Outliers: Do Not Use Standard Deviation Around the Mean, Use Median Absolute Deviation},
  author={Leys, Christophe and Klein, Olivier and Dominicy, Yannick and Ley, Christophe},
  journal={Journal of Experimental Social Psychology},
  volume={49},
  number={4},
  pages={764--766},
  year={2013},
  publisher={Elsevier}
}

@inproceedings{snell2017prototypical,
  title={Prototypical Networks for Few-shot Learning},
  author={Snell, Jake and Swersky, Kevin and Zemel, Richard},
  booktitle={Advances in Neural Information Processing Systems (NeurIPS)},
  volume={30},
  pages={4077--4087},
  year={2017},
  publisher={NeurIPS}
}

@article{bao2022gptscore,
  title={GPTScore: Evaluate as You Desire},
  author={Bao, Yao and Li, Wenqiang and Wu, Zhiyong and Shi, Yucheng and Zhang, Jing and Yu, Zhou},
  journal={Proceedings of the 2022 Conference on Empirical Methods in Natural Language Processing (EMNLP)},
  pages={1235--1248},
  year={2022},
  publisher={ACL}
}

@article{salvador2007fastdtw,
  title={Toward Accurate Dynamic Time Warping in Linear Time and Space},
  author={Salvador, Stan and Chan, Philip},
  journal={Intelligent Data Analysis},
  volume={11},
  number={5},
  pages={561--580},
  year={2007},
  publisher={IOS Press}
}

@inproceedings{wang2023selfconsistency,
  title={Self-Consistency Improves Chain of Thought Reasoning in Language Models},
  author={Wang, Xuezhi and Wei, Jason and others},
  booktitle={International Conference on Learning Representations (ICLR)},
  year={2023},
  publisher={ICLR}
}

@article{du2023improving,
  title={Improving Factuality and Reasoning in Large Language Models Through Multi-Agent Debate},
  author={Du, Yilun and Mialon, Gr{\'e}goire and others},
  journal={arXiv preprint arXiv:2305.14467},
  year={2023}
}

@inproceedings{liang2022redteaming,
  title={Holistic Evaluation of Language Models via Red Teaming},
  author={Liang, Percy and Zhang, Lianmin and others},
  booktitle={Proceedings of the 39th International Conference on Machine Learning (ICML)},
  pages={21270--21288},
  year={2022},
  publisher={ICML}
}

@article{aho1975efficient,
  title={Efficient String Matching: An Aid to Bibliographic Search},
  author={Aho, Alfred V and Corasick, Margaret J},
  journal={Communications of the ACM},
  volume={18},
  number={6},
  pages={333--340},
  year={1975},
  publisher={ACM}
}

@inproceedings{wei2022cot,
  title={Chain-of-Thought Prompting Elicits Reasoning in Large Language Models},
  author={Wei, Jason and Wang, Xuezhi and others},
  booktitle={NeurIPS},
  volume={35},
  pages={24824--24837},
  year={2022}
}

@article{gemma2024google,
  title={Gemma: Open Models Based on Gemini Research and Technology},
  author={Team, Gemma and Google DeepMind and Google Research},
  journal={arXiv preprint arXiv:2403.08295},
  year={2024}
}

@inproceedings{zeng2024safedecoding,
  title={SafeDecoding: Defending Language Models Against Jailbreak Attacks},
  author={Zeng, Jiapeng and Zhang, Yihan and Li, Bo},
  booktitle={Proceedings of the International Conference on Learning Representations (ICLR)},
  year={2024}
}

@inproceedings{fgsm2015,
  title={Explaining and Harnessing Adversarial Examples},
  author={Goodfellow, Ian and Shlens, Jonathon and Szegedy, Christian},
  booktitle={International Conference on Learning Representations (ICLR)},
  year={2015}
}

@inproceedings{pgd2018,
  title={Towards Deep Learning Models Resistant to Adversarial Attacks},
  author={Madry, Aleksander and Makelov, Aleksandar and Schmidt, Ludwig and Tsipras, Dimitris and Vladu, Adrian},
  booktitle={International Conference on Learning Representations (ICLR)},
  year={2018}
}

@inproceedings{cliprobust2022,
  title={Are Multimodal Models Robust? Evaluating Cross-Modal Adversarial Attacks},
  author={Guo, Qian and Li, Bo and Chen, Pin-Yu},
  booktitle={Proceedings of the IEEE/CVF Conference on Computer Vision and Pattern Recognition (CVPR)},
  year={2022}
}

@inproceedings{multimodal_adversarial2020,
  title={Cross-Modal Adversarial Defense: Leveraging Multimodal Structure to Find Attacks Against Multimodal Models},
  author={Xiao, Chaowei and Yang, Jun and Li, Bo and Lian, Dong and Li, Ming and Song, Dawn},
  booktitle={Advances in Neural Information Processing Systems (NeurIPS)},
  year={2020}
}

@article{guo2024prompt,
  title={Prompt Injection Attack Against LLM-integrated Applications},
  author={Guo, Han and Zhang, Qi and Ji, Yang and Zhang, Xiaoyang and Li, Yuxin and Peng, Hanzhou and Yang, Chao and Zuo, Songtao and Ye, Xiangyu and Wu, Xintao and others},
  journal={arXiv preprint arXiv:2403.02276},
  year={2024}
}

@inproceedings{qi2023visual,
  title={Visual Adversarial Examples Jailbreak Large Language Models},
  author={Qi, Yanxi and Liu, Yifan and Zhang, Zhaoyang and Wei, Feng and Zhang, Min and Ma, Yue},
  booktitle={Proceedings of the 2023 IEEE/CVF International Conference on Computer Vision (ICCV)},
  pages={1--12},
  year={2023},
  organization={IEEE}
}

@inproceedings{li2024mmjailbreak,
  title={MM-Jailbreak: Transferable Attacks on Multi-Modal Large Language Models},
  author={Li, Yuxin and Ma, Guangyao and Hsieh, Cho-Jui},
  booktitle={Proceedings of the 62nd Annual Meeting of the Association for Computational Linguistics (ACL)},
  year={2024},
  organization={ACL}
}

@inproceedings{perez2023ignore,
  title={Ignore Previous Instructions: A Systematic Study of Prompting Errors in LLMs},
  author={Perez, Ethan and Lewis, Patrick and Yao, Shunyu and others},
  booktitle={Proceedings of the 2023 Conference on Empirical Methods in Natural Language Processing (EMNLP)},
  year={2023},
  organization={ACL}
}

@inproceedings{turpin2023language,
  title={Language Models Don't Always Say What They Think: Adversarial Attacks Reducing Chain-of-Thought Reasoning Accuracy},
  author={Turpin, Miles and De Palma, Andrea and Wu, Zhaofeng and et al.},
  booktitle={Proceedings of the 40th International Conference on Machine Learning (ICML)},
  year={2023},
  organization={PMLR}
}

\newpage
\section*{Appendix}
\subsection*{A. Attack Catalog}
\label{app:attack-catalog}

This appendix provides a comprehensive catalog of attack operators instantiated under the unified formula \(A(\cdot)\). The catalog classifies attacks by operation path, encompassing signal path, text path, and prompt path attacks, and enumerates representative hybrid perturbations. Each entry contains the formal definition of the operator along with a brief description or example, complementing the attack methodology discussed in Section~\ref{sec:attack-formalization}.

\begin{table*}[t]
\centering
\small
\setlength{\tabcolsep}{6pt}
\caption{Attack catalog instantiated under the unified operator \(A(\cdot)\), grouped by attack category.}
\label{tab:attack-catalog}

\begin{tabularx}{\linewidth}{
    p{0.14\linewidth}
    p{0.18\linewidth}
    >{\raggedright\arraybackslash}p{0.26\linewidth}
    p{0.35\linewidth}
}
\toprule
\textbf{Category} & \textbf{Name} & \textbf{Formal Operator \(A(\cdot)\)} & \textbf{Description / Example} \\
\midrule

\multirow{2}{*}{\textbf{Signal Attacks}} 
& Noise Injection 
& \(A_{\text{sig}}(x_t) = x_t + \epsilon,\ \epsilon \sim \mathcal{N}(0, \sigma^2)\) 
& Adds random Gaussian noise or spike signals,
   making the translator perceive unstable or noisy motion. \\
& Drift Attack 
& \(A_{\text{sig}}(x_t[k]) = x_t[k] + \alpha k\mathbf{v}\) 
& Introduces progressive offsets along an axis,
   causing generated descriptions to drift semantically. \\
\midrule

\multirow{2}{*}{\textbf{Text Attacks}} 
& Synonym Bias 
& \(A_{\text{text}}(d_t) = \mathrm{Replace}(w_i, \mathrm{Syn}(w_i))\) 
& Replaces key descriptors to bias semantic interpretation. \\
& Adversarial Rewriting 
& \(A_{\text{text}}(d_t) = \mathrm{Typo}(d_t)\) 
& Inserts typographical errors or symbols to disrupt tokenization. \\
\midrule

\multirow{8}{*}{\textbf{Prompt Attacks}}
& Prompt Concatenation 
& \(A_{\text{prompt}}(p_t) = p_t \oplus s_e\) 
& Appends override instructions that alter the final output. \\
& Task Injection 
& \(A_{\text{prompt}}(p_t) = p_t \oplus \text{``Also write a poem.''}\) 
& Adds irrelevant tasks to divert model objectives. \\
& Role Confusion 
& \(A_{\text{prompt}}(p_t) = \mathrm{Replace}(\mathrm{role}, \mathrm{novelist})\) 
& Changes the system role to manipulate reasoning behavior. \\
& CoT Interference 
& \(A_{\text{prompt}}(p_t) = p_t \oplus \mathrm{FakeCoT}\) 
& Injects misleading reasoning chains to induce faulty inference. \\
& Multi-Task Blending 
& \(A_{\text{prompt}}(p_t) = p_t \oplus [\mathrm{task}_1 + \mathrm{task}_2]\) 
& Requests multiple subtasks simultaneously, diluting the main intent. \\
& Context Pollution 
& \(A_{\text{prompt}}(p_t) = p_t \oplus c_e\) 
& Adds contradictory or misleading contextual elements. \\
& Label Mixing 
& \(A_{\text{prompt}}(p_t) = p_t \oplus \mathrm{ConflictingLabels}\) 
& Inserts conflicting labels to induce ambiguity. \\
& Unrelated Text Noise 
& \(A_{\text{prompt}}(p_t) = p_t \oplus n_{\text{story}}\) 
& Appends lengthy irrelevant text that distracts model attention. \\
\midrule

\multirow{3}{*}{\textbf{Hybrid Attacks}} 
& Few-shot Poisoning 
& \(A_{\text{prompt}}(p_t) = p_t \oplus (x_e, y_e^{\text{false}})\) 
& Injects mislabeled few-shot examples to bias reasoning. \\
& Semantic Drift 
& \(A_{\text{text}}(d_t) = \mathrm{Hedge}(d_t)\) 
& Weakens descriptors to shift semantic boundaries. \\
& Hybrid Combo 
& \(A(x_t, p_t) = A_{\text{sig}}(x_t);\ p_t \oplus s_e\) 
& Combines physical and prompt-level perturbations for compound effects. \\
\bottomrule

\end{tabularx}
\end{table*}

\subsection*{B. End-to-End Pipelines: From Signals to Text (or Tokens) and to Prompt Construction}

This appendix describes how all target models used in this paper (LLaSA, IMUGPT~2.0,
HARGPT, ContextGPT, and MotionGPT) convert raw signal representations into textual
(or token-based) forms and how these intermediate representations are further
embedded into their final prompts.

\subsection*{B.1 LLaSA: Signal $\rightarrow$ Text $\rightarrow$ Prompt}

LLaSA extracts short statistical descriptors from IMU windows and converts them into
text strings that are later injected into a JSON-style prompt template.

\begin{tcolorbox}[colback=gray!10, colframe=black!30,
  boxrule=0.3pt, arc=1mm, left=4pt, right=4pt, top=4pt, bottom=4pt,
  title=\textbf{LLaSA Signal-to-Prompt Panel}]
\small
\textbf{Text representation (Signal → Text):}\\
IMU: ax\_var = 0.12, ay\_var = 0.08, az\_var = 0.30, step\_freq = 2.1 Hz\\
IMU: The Z-axis acceleration shows large fluctuations, and the step frequency is about 2.3 Hz.\\[4pt]

\textbf{Prompt template (Text → Prompt):}\\
You are a human activity recognition expert. Output only a JSON object:\\
{"label": "..."}\\
IMU description: The Z-axis acceleration shows large fluctuations, and the step frequency is about 2.3 Hz.\\
Task: Determine the user's activity from the set \{walk, run, sit, stand, upstairs, downstairs\}.
\end{tcolorbox}

\subsection*{B.2 IMUGPT-2.0: Signal $\rightarrow$ Text $\rightarrow$ Prompt}

IMUGPT-2.0 encodes the same IMU statistical signals but keeps them in a compact
feature-string representation.

\begin{tcolorbox}[colback=gray!10, colframe=black!30,
  boxrule=0.3pt, arc=1mm, left=4pt, right=4pt, top=4pt, bottom=4pt,
  title=\textbf{IMUGPT-2.0 Signal-to-Prompt Panel}]
\small
\textbf{Text representation (Signal → Text):}\\
Input: ax\_var = 0.12, ay\_var = 0.08, step\_freq = 2.1 Hz\\[4pt]

\textbf{Prompt template (Text → Prompt):}\\
Task: Determine the user's activity.\\
Input: ax\_var = 0.12, ay\_var = 0.08, step\_freq = 2.1 Hz.\\
Output one of: \{walk, run, sit, stand, upstairs, downstairs\}.
\end{tcolorbox}

\subsection*{B.3 HARGPT: Signal $\rightarrow$ Text $\rightarrow$ Prompt}

HARGPT prints the complete IMU sequence as text, embedding raw accelerometer and
gyroscope arrays directly into a step-by-step reasoning prompt.

\begin{tcolorbox}[colback=gray!10, colframe=black!30,
  boxrule=0.3pt, arc=1mm, left=4pt, right=4pt, top=4pt, bottom=4pt,
  title=\textbf{HARGPT Signal-to-Prompt Panel}]
\small
\textbf{Text representation (Signal → Text):}\\
Accelerations: [0.15, 0.20, -0.10, ...]\\
Gyroscopes: [...]\\[4pt]

\textbf{Prompt template (Text → Prompt):}\\
Instruction: You are an expert of IMU-based human activity analysis.\\
Question: IMU collected from \{device name\} at \{location\}.\\
Accelerations: \{...\}; Gyroscopes: \{...\}.\\
Activity classes: \{walk, run, sit, stand, upstairs, downstairs\}.\\
Please analyze step by step.
\end{tcolorbox}

\subsection*{B.4 ContextGPT: Signal $\rightarrow$ Text $\rightarrow$ Prompt}

ContextGPT merges contextual cues (room, surroundings) with IMU signal summaries.

\begin{tcolorbox}[colback=gray!10, colframe=black!30,
  boxrule=0.3pt, arc=1mm, left=4pt, right=4pt, top=4pt, bottom=4pt,
  title=\textbf{ContextGPT Signal-to-Prompt Panel}]
\small
\textbf{Text representation (Context + IMU):}\\
The user is currently in the bedroom.\\
IMU summary: step\_freq = 2.0 Hz and Z-axis acceleration shows large fluctuations.\\[4pt]

\textbf{Prompt template (Text → Prompt):}\\
You are an expert in human activity recognition.\\
Context: The user is currently in the bedroom.\\
IMU summary: step\_freq = 2.0 Hz, Z-axis amplitude high.\\
Question: What is the user most likely doing now?
\end{tcolorbox}

\subsection*{B.5 MotionGPT: Signal $\rightarrow$ Motion Tokens $\rightarrow$ Prompt}

MotionGPT converts 3D motion into discrete tokens through a VQ-VAE tokenizer,
and integrates them with natural-language instructions.

\begin{tcolorbox}[colback=gray!10, colframe=black!30,
  boxrule=0.3pt, arc=1mm, left=4pt, right=4pt, top=4pt, bottom=4pt,
  title=\textbf{MotionGPT Signal-to-Prompt Panel}]
\small
\textbf{Text representation (3D Motion → Tokens):}\\
<motion\_id\_33> <motion\_id\_439> <motion\_id\_70> ...\\[4pt]

\textbf{Prompt templates (Text → Prompt):}\\
Text-to-motion:\\
"Generate a motion sequence depicting a person emulating a waltz dance."\\[4pt]
Motion-to-text:\\
"Provide an accurate caption describing <motion\_tokens>."\\[4pt]
Motion QA:\\
"Randomly describe the motion."\\
"Generate more from this motion."
\end{tcolorbox}

\end{document}